\documentclass[12pt]{article}
\usepackage{amssymb,amsmath,epsfig}
\usepackage{graphicx}
\allowdisplaybreaks

\begin{document}
\title{\textbf{Stability and Physical Properties of Compact Stars Beyond
Einstein Gravity}}
\author{M. Sharif$^{1,2}$\thanks {msharif.math@pu.edu.pk}
~and Adeeba Arooj$^1$\thanks{aarooj933@gmail.com}\\
$^1$Department of Mathematics and Statistics, The University of Lahore,\\
1-KM Defence Road Lahore-54000, Pakistan.\\
$^2$Research Center of Astrophysics and Cosmology, Khazar
University,\\ Baku, AZ1096, 41 Mehseti Street, Azerbaijan.}
\date{}
\maketitle

\begin{abstract}
This manuscript discusses feasible features of anisotropic celestial
sphere within the framework of $f(\mathbb{Q},\mathcal{L}_{m})$
gravity, where $\mathbb{Q}$ represents non-metricity scalar and
$\mathcal{L}_{m}$ is the matter Lagrangian. The geometric
configuration of static spherical symmetric structure is examined
using a specific non-singular solution (Krori-Barua solution). A
particular model of this theory is considered to derive explicit
field equations. The Darmois matching conditions are used to
evaluate unknown constants in the metric coefficients. To verify
plausible existence of compact objects in this gravitational
framework, we analyze their fundamental physical properties
including fluid parameters, gradients, surface redshift, mass-radius
relation, anisotropy measure, compactness factor, energy conditions
and equations of state. The stability of the considered stellar
objects is verified by adiabatic index and sound speed. Our results
demonstrate that all required physical conditions are satisfied,
confirming the existence of physically stable anisotropic celestial
objects within this modified gravity.
\end{abstract}
\textbf{Keywords:} Celestial objects; Stability; Energy constraints;
$f(\mathbb{Q},\mathcal{L}_{m})$ theory. \\
\textbf{PACS:} 97.10.Ex; 97.10.Tk; 95.55.-n; 98.58.M; 75.10.-b;
97.60.Jd; 98.52.-b.

\section{Introduction}

Our galaxy is a vast collection of stars and their planetary
systems, all gravitationally allied together. Numerous observations
have been made regarding the shapes and properties of galaxies, as
well as the groups in which they are found. Stars within a galaxy
sharing identical mass and formed simultaneously under similar
conditions, exhibit identical observed properties. They utilize
gravitational force to balance the outward pressure generated by
nuclear fusion \cite{1}. However, these massive stellar bodies
eventually collapse, forming celestial objects that vary in mass,
size and characteristics. Among all, the most fascinating celestial
objects in the history of the universe are black holes, quark stars,
black dwarfs, neutron stars and white dwarfs.

A celestial black hole is formed from the collapse of star with a
mass ranging from five to several tens of solar mass ($m_{\odot}$).
The concept of strong gravitational black holes was first proposed
by John Michell in 1783. This idea was further developed within the
framework of general relativity by Karl Schwarzschild around 1916.
Due to their intriguing nature, researchers embraced the
relativistic effects of black holes such as gravitational time
dilation. Black holes also became a recurring element in science
fiction, particularly as a mechanism for space travel. Within this
framework, a black hole is often theorized to connect a white hole,
acting as a gateway to distant region of space potentially far from
the point of entry. More speculatively, some representations suggest
that the exit point might exist in a different time region, enabling
time travel or providing a passage to an entirely separate universe
\cite{1a}-\cite{1aaa}.

A celestial neutron star created by supernovae \cite{2} has mass of
upto 3.0 $m_{\odot}$. These stars emit intense radiation
particularly X and gamma-rays, due to the strong magnetic fields at
their poles \cite{3}. Additionally, neutron stars significantly
influence the production of gravitational waves and synthesis of
heavy elements in the universe. Their remarkable characteristics
such as high density, extremely strong magnetic field and rapid
rotation have attracted the interest of numerous researchers
\cite{4}-\cite{12}. The celestial white dwarf is a hot dense
earth-sized stellar remnant, primarily composed of
electron-degenerate matter. This final evolutionary material include
quantum mechanical state where atomic nuclei (carbon and oxygen) are
suspended in a sea of electrons. A white dwarf is prevented from
collapse under its own immense gravity not by thermal pressure (like
a main-sequence star) but by electron degeneracy pressure.

Modified gravitational theories have emerged as viable alternatives
to general relativity, offering potential explanation for various
cosmological and astrophysical phenomena without invoking dark
matter or dark energy \cite{13}. The exploration of such theories is
largely motivated by challenges posed by the cosmological constant,
which is characterized by a significant discrepancy between the
vacuum energy density predicted by quantum field theory and the
value inferred from cosmological observations \cite{14}. A novel
geometric framework has been developed to accurately replicate
gravitational dynamics while circumventing the cosmological constant
problem. This framework employs a linear, causal and nonlocal
functional relationship between the metric tensor, radiations and
the energy-momentum tensor. To overcome existing theoretical
obstacles and unravel secrets of the universe, scientists have
developed a variety of modified gravitational theories
\cite{15}-\cite{20}. Among these, $f(R)$ gravity has garnered
significant attention due to its relative simplicity and ability to
produce a wide range of cosmological behaviors. In $f(R)$ gravity,
the Ricci scalar in the Einstein-Hilbert action is replaced by a
generic function of $R$, thereby modifying the gravitational
dynamics. This modification can lead to accelerated expansion even
in a matter-dominated universe \cite{21,22}.

Through the lens of connection geometry, Weyl \cite{23} developed a
more comprehensive geometric framework incorporating non-metricity
with a goal of unifying the fundamental forces of nature. This
approach focuses on the behavior of a conformal factor rather than
the preservation of vector length during parallel transport. By
asserting that the covariant derivative of the metric tensor is
non-zero, Weyl's theory introduces the concept of non-metricity.
Non-Riemannian geometries extend our understanding of spacetime by
incorporating torsion and non-metricity. Torsion serves as the
representation of gravitational interaction in teleparallel gravity,
expressed through the action $\int\sqrt{-g} T d^4x$ (T represent
torsion) \cite{24}. Non-metricity fulfills a similar role in
symmetric teleparallel gravity, described by $\int \sqrt{-g}
\mathbb{Q}  d^4x$ ($\mathbb{Q}$ represent non-metricity) \cite{25}.
Further extending this framework, a joint presence of non-metricity
and the trace of energy-momentum tensor characterize gravitational
interaction in extended symmetric teleparallel gravity, given by the
action $\int\sqrt{-g} f(\mathbb{Q},\mathbb{T})  d^4x$ ($\mathbb{T}$
represent trace of energy-momentum) \cite{26}-\cite{30}.

Ayush Hazarika generalized coincident general relativity and
symmetric teleparallel gravity to develop a modified gravitational
theory named as the $f(\mathbb{Q},\mathcal{L}_{m})$ theory
\cite{31}. The field equations for this theory were developed by
generalizing the field equations of $f(\mathbb{Q})$ gravity. This
theory employs a unique coupling of matter and geometry to address
key challenges in gravitational physics. The incorporation of
non-metricity broadens a scope for viable cosmological dynamics,
while the explicit dependence on lagrangian density directly encodes
how matter influences gravity. A cosmological implications of the
$f(\mathbb{Q},\mathcal{L}_{m})$ theory were examined for a flat,
homogeneous and isotropic FLRW geometry. Motivated by these
theoretical advantages, researchers have actively investigated the
theory's implications. Myrzakulov et al examined late-time cosmic
expansion and energy conditions in this theory \cite{32}-\cite{34}.
Samaddar et al investigated the approach to baryogenesis and its
cosmological implications within the same framework \cite{35}. Gul
et al \cite{36} examined the ghost dark energy model in this
modified theory. Sharif et al \cite{37} investigated viable
cosmological bounce solutions in the same gravity.

The $f(\mathbb{Q},\mathcal{L}_{m})$ theory introduces a specific
connection of geometry and matter, capturing considerable attention
among researchers due to its profound implications in gravitational
physics. The exploration of this theory is motivated by a desire to
understand its theoretical implications and significance in
astrophysical as well as cosmological contexts. In
$f(\mathbb{Q},\mathcal{L}_{m})$ theory, the inclusion of
non-metricity and the presence of matter source enable a more
detailed depiction of gravitational interactions. Non-metricity
represents the deviation from the Levi-Civita connection, which is
the connection compatible with the metric tensor in GR. By
incorporating non-metricity into the gravitational action, this
theory introduces additional degrees of freedom, leading to new
gravitational dynamics and cosmological solutions. Furthermore, the
dependence on the matter-Lagrangian allows this theory to capture
the effects of matter content on the gravitational field. The
recently discovered $f(\mathbb{Q},\mathcal{L}_{m})$ theory highlight
the diverse avenues of research in reshaping our understanding of
cosmic evolution. Myrzakulov et al investigated cosmology within the
framework of modified $f(\mathbb{Q},\mathcal{L}_{m})$ gravity, while
using the non-linear model \cite{01}. Rana et al investigated the
influence of viscosity on the evolution of the cosmos within the
framework of $f(\mathbb{Q},\mathcal{L}_{m})$ gravity \cite{02}.
Thus, the motivation behind this theory is to consider a more
comprehensive framework for gravitational physics and cosmology,
capable of addressing current observational discrepancies and
offering insights into fundamental questions about the nature of
gravity and the universe.

In the dense interiors of compact stars, the radial and tangential
pressure of fluid are not expected to be equal. Anisotropic
configuration naturally arises at supranuclear densities due to a
variety of microphysical mechanisms and therefore constitutes a
realistic physical condition for modeling CSs. The interactions
between particles become highly directional, new degrees of freedom
may appear and the fluid may acquire shear stresses. As a result,
the effective radial and tangential pressure differ, producing
pressure anisotropy. Several theoretical studies and microphysical
models support the emergence of anisotropy in CSs. Herrera and
Santos discussed possible causes for the appearance of local
anisotropy (principal stresses unequal) in self-gravitating systems
and present its main consequences by considering both Newtonian and
general relativistic examples \cite{03}. Herrera investigated the
conditions for the (in)stability of the isotropic pressure condition
in collapsing spherically symmetric, dissipative fluid distributions
\cite{04}. Hence, modeling the matter content through an anisotropic
fluid is both realistic and necessary for constructing viable
stellar solutions in modified theories of gravity.

Cosmic expansion laid a great interest in alternative gravitational
theories, which have been developed from Einstein's gravitational
theory. Mustafa studied the realistic nature and viability of
wormhole solutions in the Ricci inverse gravity \cite{05}. Kavya et
al explored wormhole solutions in curvature-matter coupling gravity
supported by non-commutative geometry and conformal symmetry
\cite{06}. Mustafa et al discussed stable thin-shell around
wormholes within string cloud and quintessential field \cite{07}.
Mustafa et studied the dynamical characteristics of an anisotropic
CSs with spherical symmetry \cite{08}. Maurya studied charged CSs in
extended symmetric teleparallel gravity and discussed the influence
of charge and coupling parameters on the stellar configuration
\cite{09}. Amin et al modeled CSs in the framework of modified
teleparallel gravity \cite{011}. Khatoon et al discussed impact of
geometric deformation and anisotropy on CSs in modified rainbow
gravity \cite{012}. Mustafa et al investigate anisotropic fluid
spheres admitting Karmarkar condition in modified gravity
\cite{013}. Murtaza et al introduced charged black hole \cite{014}.
Zubair et al examined non-commutative wormholes admitting conformal
motion involving minimal coupling \cite{015}. Bouali et al undergoes
cosmological tests of parametrization in $f(\mathbb{Q})$ theory
\cite{016}.

The recently discovered $f(\mathbb{Q},\mathcal{L}_{m})$ gravity
represents a novel synthesis that extends existing models by
combining the non-Riemannian geometric foundation of $f(\mathbb{Q})$
gravity with a specific type of geometry-matter coupling explored in
Riemannian theories. Its distinction lies in the dual nature of its
generalization. Firstly, it differs fundamentally from
$f(R,\mathcal{L}_{m})$ theory by changing the very geometric
description of gravity which operates within the standard Riemannian
geometry of GR, using curvature $(R)$ as its primary geometric
variable. While $f(\mathbb{Q},\mathcal{L}_{m})$ gravity is built
upon symmetric teleparallel geometry, where gravity is manifested
through non-metricity $(\mathbb{Q})$ in a flat, torsion-free
spacetime. This grants $f(\mathbb{Q},\mathcal{L}_{m})$ gravity the
mathematical advantage of second-order field equations, unlike the
higher-order equations often found in $f(R)$ theory.

The $f(\mathbb{Q},\mathcal{L}_{m})$ gravity extends the simpler
$f(\mathbb{Q})$ gravity by introducing a non-minimal coupling
between $(\mathbb{Q})$ and the matter fields embodied by the
matter-Lagrangian $(\mathcal{L}_{m})$. This is a different and more
fundamental coupling than the one found in
$f(\mathbb{Q},\mathbb{T})$ gravity. Although, the
$f(\mathbb{Q},\mathbb{T})$ couples geometry to the trace of the
energy-momentum tensor $(\mathbb{T})$, a single scalar derived from
the macroscopic properties of matter but the
$f(\mathbb{Q},\mathcal{L}_{m})$ gravity coupling is to the full
Lagrangian density itself. This means the interaction is sensitive
to the microscopic structure and theoretical formulation of the
matter fields, leading to more nuanced and potentially richer
dynamics. Consequently, $f(\mathbb{Q},\mathcal{L}_{m})$ gravity is
not merely a competitor to these models but a more comprehensive
framework that leverages the benefits of a non-Riemannian geometry
while embedding a deep, Lagrangian-based coupling between the
structure of spacetime and the fundamental nature of matter.

The aforementioned study persuades us to explore viable
characteristics of anisotropic celestial objects within the
framework of $f(\mathbb{Q},\mathcal{L}_{m})$ gravity. In this paper,
section \emph{2} outlines the fundamental of this theory and derives
explicit expressions for the field equations. In section \emph{3},
various physical quantities including fluid parameters, their
gradients, anisotropy, energy conditions, compactness, mass,
equation of state and redshift are explored to assess viable
anisotropic celestial objects. Section \emph{4} analyzes stability
of the anisotropic compact objects. Final section summarizes our
results, by confirming stability of anisotropic celestial objects
within this modified gravity.

\section{Field Equations in $f(\mathbb{Q},\mathcal{L}_{m})$ Gravity}

The mathematical basis of general relativity relies on Riemannian
geometry, where parallel transport preserves both direction and
length of vector along a closed path. Weyl \cite{23} presented
generalization in which a vector's magnitude and orientation may
vary during parallel transport. This introduces a novel vector field
$(Y^{\chi})$, which determines the geometric structure of Weyl
spacetime. In Weyl's theory, the length of a vector changes as
$\delta\ell= \ell Y_{\chi} \delta x^ {\chi}$, where $\ell$ is the
size of vector and $\delta x^{\chi}$ defines infinitesimal path.
This implies that the change in vector's length depends on
connection coefficients, initial length and displacement path. A
change in vector's length is given by $\delta\ell=
\ell\Upsilon_{\chi\tau}\delta h^{\chi\tau}$, where $\delta
h^{\chi\tau}$ is the area element and
\begin{equation}\label{1a}
\Upsilon_{\chi\tau}=\nabla_{\tau}Y_{\chi}-\nabla_{\chi}Y_{\tau}.
\end{equation}
A spatial transformation $\hat{\ell}=\upsilon(x)\ell$ alters the
field equation $\hat{Y}_{\chi}$ to $\hat{Y}_{\chi}=Y_{\chi}+
(\ln\upsilon)_{\chi}$, while conformal transformations adjust the
components of the metric tensor as
$\hat{g}_{\chi\tau}=\upsilon^{2}g_{\chi\tau}$ and
$\hat{g}^{\chi\tau}= \upsilon^{-2}g^{\chi\tau}$. Weyl geometry
encompasses various characteristics, defined as
\begin{equation}\label{2a}
{\hat{\Gamma}}^{\upsilon}_{~\chi\tau}=\Gamma^{\upsilon}_{~\chi\tau}
+g_{\chi\tau}Y^{\upsilon}-\delta^{\upsilon}_{~\chi}Y_{\tau}-
\delta^{\upsilon}_{~\tau}Y_{\chi}.
\end{equation}
The Christoffel symbol $(\Gamma^{\upsilon}_{~\chi\tau})$ is the
affine connection in this framework. By requiring symmetry in the
modified connection coefficient
$(\hat{\Gamma}^{\upsilon}_{~\chi\tau})$, we can properly define a
gauge-covariant derivative. This derivative operator yields the Weyl
curvature tensor through standard geometric construction as
\begin{equation}\label{3a}
\hat{S}_{\chi\tau\upsilon\gamma}=\hat{S}
_{(\chi\tau)\upsilon\gamma}+\hat{S}_{[\chi\tau]\upsilon\gamma},
\end{equation}
where
\begin{equation}\nonumber
\hat{S}_{[\chi\tau]\upsilon\gamma}=S
_{\chi\tau\upsilon\gamma}+2\nabla_{\upsilon}Y_{[\chi
g_{\tau}]\gamma}+2\nabla_{\gamma}Y_{[\tau
g_{\chi}]\upsilon}+2Y_{\upsilon}Y_{[\chi
g_{\tau}]\upsilon}+2Y_{\gamma}Y_{[\tau
g_{\chi}]\upsilon}-2Y^{2}g_{\upsilon[\chi g_{\tau}]\gamma}.
\end{equation}
Applying the first contraction operation to the Weyl tensor, we have
\begin{eqnarray}\label{5a}
\hat{S}^{\chi}_{~\tau}&=&S^{\chi}_{~\tau}
+2Y^{\chi}Y_{\tau}+3\nabla_{\tau}Y^{\chi}-\nabla_{\chi}Y^{\tau}
+g^{\chi}_{~\tau}(\nabla_{\upsilon}Y^{\upsilon}
-2Y_{\upsilon}Y^{\upsilon}),\\\label{6a}
\hat{S}&=&\bar{S}^{\upsilon}_{~\upsilon}=
S+6(\nabla_{\chi}Y^{\chi}-Y_{\chi}Y^{\chi}).
\end{eqnarray}

Weyl-Cartan (WC) geometries extend beyond the Riemannian and Weyl
frameworks by incorporating non-zero torsion. The WC geometry uses a
symmetric metric tensor for length measurement and an asymmetric
connection that regulates parallel transport as
$d\varsigma^{\chi}=-\varsigma^{\upsilon}
\hat{\Gamma}^{\chi}_{~\upsilon\tau}dx^{\tau}$.
The connection $(\tilde{\Gamma}^{\upsilon}_{~\chi\tau})$,
disformation tensor $(\Omega^{\upsilon}_{~\chi\tau})$ and contortion
tensor $(\Psi^{\upsilon}_{~\chi\tau})$ in this framework are given
as
\begin{eqnarray}\label{7a}
\tilde{\Gamma}^{\upsilon}_{~\chi\tau}&=&{\Gamma}^{\upsilon}_{~\chi\tau}
+\Psi^{\upsilon}_{~\chi\tau}+\Omega^{\upsilon}_{~\chi\tau},
\\\label{8a}
\Omega^{\upsilon}_{~\chi\tau}&=&\frac{1}{2}g^{\upsilon\gamma}
(\mathbb{Q}_{\chi\tau\gamma} +\mathbb{Q}_{\chi\tau\gamma}-\mathbb{Q}_{\gamma\chi\tau}),
\\\label{9a}
\Psi^{\upsilon}_{~\chi\tau}&=&\tilde{\Gamma}^{\upsilon}_{~[\chi\tau]}
+g^{\upsilon\gamma}g_{\chi\varepsilon}
\tilde{\Gamma}^{\varepsilon}_{~[\tau\gamma]}+g^{\upsilon\gamma}
g_{\tau\varepsilon}\tilde{\Gamma}^{\varepsilon}_{~[\chi\gamma]},
\end{eqnarray}
where
\begin{equation}\label{10a}
\mathbb{Q}_{\gamma\chi\tau}=\nabla_{\gamma}g_{\chi\tau}
=-g_{\chi\tau,\gamma}+g_{\tau\varepsilon}\hat{\Gamma}^{\varepsilon}_{~\chi\gamma}
+g_{\varepsilon\chi}\hat{\Gamma}^{\varepsilon}_{~\tau\gamma}.
\end{equation}
The WC framework reduces to the standard Weyl geometry when the
torsion tensor vanishes as demonstrated by Eqs.(\ref{2a}) and
(\ref{7a}), where $\mathbb{Q}_{\upsilon\chi\tau} =
-2g_{\chi\tau}\Psi_{\upsilon}$. Therefore, Eq.(\ref{7a}) becomes
\begin{equation}\label{11a}
\tilde{\Gamma}^{\upsilon}_{~\chi\tau}={\Gamma}^{\upsilon}_{~\chi\tau}
+g_{\chi\tau}\Psi^{\upsilon}
-\delta^{\upsilon}_{~\chi}\Psi_{\tau}-\delta^{\upsilon}_{~\tau}\Psi_{\chi}
+\Psi^{\upsilon}_{~\chi\tau},
\end{equation}
with
\begin{equation}\label{12a}
\Psi^{\upsilon}_{~\chi\tau}=T^{\upsilon}_{~\chi\tau}-g^{\upsilon\gamma}
g_{\varepsilon\chi}T^{\varepsilon}_{~\gamma\tau}-g^{\upsilon\gamma}
g_{\varepsilon\tau}T^{\varepsilon}_{~\gamma\chi}.
\end{equation}
The WC torsion is given by
\begin{equation}\label{13a}
T^{\upsilon}_{~\chi\tau}=\frac{1}{2}(\tilde{\Gamma}^{\upsilon}
_{~\chi\tau}-\tilde{\Gamma}^{\upsilon}_{~\tau\chi}).
\end{equation}
The WC curvature tensor can be defined in terms of connection
coefficients as
\begin{equation}\label{14a}
\tilde{S}^{\upsilon}_{~\chi\tau\gamma}=\tilde{\Gamma}^{\upsilon}
_{~\chi\gamma,\tau}-\tilde{\Gamma}^{\upsilon}_{~\chi\tau,\gamma}+\tilde{\Gamma}
^{\varepsilon}_{~\chi\gamma}
\tilde{\Gamma}^{\upsilon}_{~\varepsilon\tau}-\tilde{\Gamma}
^{\varepsilon}_{~\chi\tau}
\tilde{\Gamma}^{\upsilon}_{~\varepsilon\gamma}.
\end{equation}
The WC scalar is obtained as
\begin{eqnarray}\nonumber
\tilde{S}&=&\tilde{S}^{\chi\tau}_{~~\chi\tau}
=S+6\nabla_{\tau}\Psi^{\tau}-4\nabla_{\tau}
T^{\tau}-6\Psi_{\tau}\Psi^{\tau} +8\beta
_{\tau}T^{\tau}+T^{\chi\upsilon\tau}T_{\chi\upsilon\tau}
\\\label{15a}
&+&2T^{\chi\upsilon\tau}T_{\tau\upsilon\chi}-4T^{\tau}T_{\tau}.
\end{eqnarray}

By omitting boundary terms in the Ricci scalar, the gravitational
action can be written as
\begin{equation}\label{16a}
S=\frac{1}{2\kappa} \int
g^{\chi\tau}(\Gamma^{\upsilon}_{~\gamma\chi}\Gamma^{\gamma}_{~\upsilon\tau}
-\Gamma^{\upsilon}_{~\gamma\upsilon}\Gamma^{\gamma}_{~\chi\tau})\sqrt{-g}
d^{4}x.
\end{equation}
For a symmetric connection, we have
\begin{equation}\label{17a}
\Gamma^{\upsilon}_{~\chi\tau}=-\mathcal{L}^{\upsilon}_{~\chi\tau}.
\end{equation}
Thus, Eq.\eqref{16a} becomes
\begin{equation}\label{18a}
S=-\frac{1}{2\kappa} \int
g^{\chi\tau}(\mathcal{L}^{\upsilon}_{~\gamma\chi}\mathcal{L}^{\gamma}_{~\upsilon\tau}
-\mathcal{L}^{\upsilon}_{~\gamma\upsilon}\mathcal{L}^{\gamma}_{~\chi\tau})\sqrt{-g}
d^{4}x,
\end{equation}
where
\begin{equation}\label{19a}
\mathbb{Q}\equiv-g^{\chi\tau}(\mathcal{L}^{\upsilon}_{~\gamma\chi}\mathcal{L}^{\gamma}_{~\upsilon\tau}
-\mathcal{L}^{\upsilon}_{~\tau\upsilon}\mathcal{L}^{\tau}_{~\chi\tau}),
\end{equation}
and
\begin{equation}\label{20a}
\mathcal{L}^{\upsilon}_{~\chi\tau}\equiv-\frac{1}{2}g^{\upsilon\gamma}
(\nabla_{\tau}g_{\chi\gamma}+\nabla_{\chi}g_{\gamma \tau}
-\nabla_{\gamma}g_{\chi\tau}).
\end{equation}
Substituting a generic function for the non-metricity scalar in
Eq.(\ref{18a}), we have
\begin{equation}\label{21a}
S=\frac{1}{2\kappa}\int f(\mathbb{Q})\sqrt{-g}d^{4}x.
\end{equation}
Coupling of this action with matter-Lagrangian yields \cite{31}
\begin{equation}\label{22a}
S=\frac{1}{2\kappa}\int f(\mathbb{Q},\mathcal{L}_{m}) \sqrt{-g}d^{4}x.
\end{equation}
The superpotential is given by
\begin{equation}\label{23a}
\mathcal{P}^{\upsilon}_{~\chi\tau}=-\frac{1}{2}\mathcal{L}^{\upsilon}_{~\chi\tau}
+\frac{1}{4}(\mathbb{Q}^{\upsilon}
-\tilde{\mathbb{Q}}^{\upsilon})g_{\chi\tau}- \frac{1}{4} \delta
^{\upsilon} _{~[\chi \mathbb{Q}_{\tau}]}.
\end{equation}
The non-metricity relation (provided in Appendix $\textbf{A}$) is
\begin{equation}\label{25a}
\mathbb{Q}=-\mathbb{Q}_{\upsilon\chi\tau}\mathcal{P}
^{\upsilon\chi\tau}=-\frac{1}{4} (-\mathbb{Q}^{\upsilon\chi\tau}
\mathbb{Q}_{\upsilon\chi\tau}+2\mathbb{Q}^{\upsilon\chi\tau}
\mathbb{Q}_{\tau\upsilon\chi}
-2\mathbb{Q}^{\upsilon}\tilde{\mathbb{Q}}_{\upsilon}+\mathbb{Q}
^{\upsilon}\mathbb{Q}_{\upsilon}).
\end{equation}
The variation of Eq.(\ref{22a}) corresponding to the metric tensor
yields
\begin{equation}\label{26a}
\delta S=\frac{1}{2}\int\delta [f(\mathbb{Q},\mathcal{L}_{m})
\sqrt{-g}]d^{4}x =\frac{1}{2}\int(f\delta\sqrt{-g}
+(f_{\mathbb{Q}}\delta \mathbb{Q}+f_{\mathcal{L}_{m}}\delta
\mathcal{L}_{m})\sqrt{-g}d^{4}x.
\end{equation}
Moreover, we define
\begin{eqnarray}\label{27a}
\mathbb{T}_{\chi\tau}=-\frac{2}{\sqrt{-g}}\frac{\delta(\sqrt{-g}\mathcal{L}_{m})}{\delta
g^{\chi\tau}}=g_{\chi\tau}\mathcal{L}_{m}-2\frac{\partial
\mathcal{L}_{m}}{\partial g^{\chi\tau}}.
\end{eqnarray}
The variation of $\mathbb{Q}$ is given in Appendix \textbf{B}. The
variation of determinant of the metric tensor is given by
\begin{equation}\label{27aa}
\delta\sqrt{-g}=-\frac{1}{2}\sqrt{-g}g_{\chi\tau}\delta
g^{\chi\tau}.
\end{equation}
Using variation of $\delta \mathbb{Q}$, it follows that
\begin{eqnarray}\nonumber
\delta S&=&\frac{-1}{2}\int f g_{\chi\tau}\sqrt{-g}\delta
g^{\chi\tau}
\\\nonumber
&-&f_{\mathbb{Q}}\sqrt{-g}(\mathcal{P}_{\chi\upsilon\gamma}
\mathbb{Q}_{\tau}^{~\upsilon\gamma}-2\mathbb{Q}^{\upsilon\gamma}_{~~\chi}
\mathcal{P}_{\upsilon\gamma\tau}) \delta
g^{\chi\tau}+2f_{\mathbb{Q}}\sqrt{-g}
\mathcal{P}_{\upsilon\chi\tau}\nabla^{\upsilon} \delta g^{\chi\tau}
\\\label{28a}
&+&\frac{1}{2}f_{\mathcal{L}_{m}}(g_{\chi\tau}\mathcal{L}_{m}-\mathbb{T}_{\chi\tau})\sqrt{-g}
\delta g^{\chi\tau}d^ {4}x.
\end{eqnarray}

The constant conventions, specifically $G=c=1$ is adopted in
cosmological studies to simplify the field equations. Setting $c=1$
corresponds to working in natural units, in which the speed of light
is normalized. This choice is particularly convenient for our
stability analysis based on the squared sound speed where the
physical requirement $0\leq u^2\leq1$ ensures that perturbations
propagate subliminally, i.e., no physical signal or particle exceeds
the speed of light. Therefore, taking $c=1$ makes the condition on
stability equivalent to demanding causality in the perturbation
dynamics. Similarly, setting $G=1$ belongs to the class of
geometrized units in which Newton's gravitational constant $G$ is
absorbed into the spacetime geometry. Here, $G$ measures the
strength of the interaction between matter and geometry. By choosing
$G=1$, the modified field equations of
$f(\mathbb{Q},\mathcal{L}_{m})$ gravity attain a simplified form
without altering the physical content. This normalization re-scales
the equations so that gravitational quantities and matter quantities
appear with consistent dimensions, allowing for cleaner analytical
expressions in cosmological reconstruction.

The corresponding field equations after simplification are
\begin{eqnarray}\nonumber
\frac{1}{2}f_{\mathcal{L}_{m}}(g_{\chi\tau}\mathcal{L}_{m}-\mathbb{T}_{\chi\tau})&=&
\frac{2}{\sqrt{-g}} \nabla_{\upsilon} (f_{\mathbb{Q}}\sqrt{-g}
\mathcal{P}^{\upsilon}_{~\chi\tau})+ \frac{1}{2}fg_{\chi\tau}
\\\label{29a}
&+&f_{\mathbb{Q}} (\mathcal{P}_{\chi\upsilon\gamma}
\mathbb{Q}_{\tau}^{~\upsilon\gamma} -2\mathbb{Q}^{\upsilon\gamma}_{~~\chi}
\mathcal{P}_{\upsilon\gamma\tau}),
\end{eqnarray}
where $f_{\mathcal{L}_{m}}=\frac{\partial f}{\partial
\mathcal{L}_{m}}$ and $f_{\mathbb{Q}}=\frac{\partial f}{\partial
\mathbb{Q}}$. The $f(\mathbb{Q},\mathcal{L}_{m})$ field equations
provide important results regarding the behavior of gravity. We
consider a static spherical spacetime with metric signatures
$(+,-,-,-)$ as
\begin{equation}\label{30a}
ds^{2}=e^{\nu(r)}dt^{2}-e^{\iota(r)}dr^{2}-r^{2}d \theta^{2}-r^{2}
\sin^{2}\theta d\phi^{2}.
\end{equation}
In order to describe matter distribution of spacetime, we consider
matter-Lagrangian as $\mathcal{L}_{m}=p_{r}$ \cite{38}. The momentum
tensor describe distribution of matter and energy within a system
and its parameters determine physical properties that govern
dynamics of the system. We examine an anisotropic matter
distribution characterized by a 4-velocity ($\mathbb{U}_{\tau}$) and
4-vector ($\mathbb{V}_{\tau}$) of the fluid as
\begin{equation}\label{31a}
\mathbb{T}_{\tau\chi}=\mathbb{U}_{\tau}\mathbb{U}_{\chi}\varrho
+\mathbb{V}_{\tau}\mathbb{V}_{\chi}p_{r}-p_{t} g_{\tau\chi}+
\mathbb{U}_{\tau}\mathbb{U}_{\chi}p_{t}
-\mathbb{V}_{\tau}\mathbb{V}_{\chi}p_{t}.
\end{equation}

The choice, $\mathcal{L}_{m}=p_r$, yields a stress-energy tensor of
the form $\mathbb{T}_{\tau\chi}$ that maintains compatibility with
the standard conservation equation in the GR limit. Moreover, this
choice avoids the introduction of extra dynamical degrees of freedom
associated with derivatives of the energy density. With
$\mathcal{L}_{m}=p_r$, the variation $\frac{\partial
\mathcal{L}_{m}}{\partial g^{\chi\tau}}$ depends only on metric
components through the pressure, simplifying the matter coupling
terms in the field equations. As a result, the modified gravity
contributions $(f_{\mathcal{L}_{m}}=\frac{\partial f}{\partial
\mathcal{L}_{m}},f_{\mathbb{Q}}=\frac{\partial f}{\partial
\mathbb{Q}})$ remain algebraically tractable, enabling closed-form
relations for effective density and pressures. Additionally, radial
pressure typically decreases from the center to the boundary, this
choice yields regular behavior of the non-minimal coupling terms
throughout the CSs. The choice of $L_{m}=p_{r}$ is strongly
supported by the literature \cite{58}-\cite{60} to yield stable and
viable astrophysical models.

The corresponding field equations take the form (Details are given
in Appendix \textbf{C})
\begin{eqnarray}\nonumber
\frac{1}{2}f_{\mathcal{L}_{m}} (\varrho- \mathcal{L}_{m})&=&\frac{1}{2
r^{2}e^{\iota}}\bigg(2r
f_{\mathbb{Q}\mathbb{Q}}\mathbb{Q}'(-1+e^{\iota})+
f_{\mathbb{Q}}((2+r \nu')(e^{\iota}-1)
\\\label{32a}
&+& r\iota'(1+e^{\iota}))\bigg)+\frac{f}{2},
\\\nonumber
\frac{1}{2}f_{\mathcal{L}_{m}}(p_{r}-\mathcal{L}_{m})&=&\frac{-1}{2
r^{2}e^{\iota}}\bigg(2r
f_{\mathbb{Q}\mathbb{Q}}\mathbb{Q}'(e^{\iota}-1)
+(e^{\iota}-1)f_{\mathbb{Q}}(2+r\nu'+r\iota')
\\\label{33a}
&-&2r\nu')\bigg)-\frac{f}{2},
\\\nonumber \frac{1}{2}f_{\mathcal{L}_{m}}(p_{t}-\mathcal{L}_{m})&=&
\frac{-1}{4re^{\iota}}\bigg(-2rf_{\mathbb{Q}\mathbb{Q}}\mathbb{Q}'\nu'+
f_{\mathbb{Q}}(2\nu'(e^{\iota}-2)-r\nu'^{2}
+\iota'(2e^{\iota}
\\\label{34a}
&+&r\nu')-2r\nu'')\bigg)-\frac{f}{2}.
\end{eqnarray}
Above equations contain derivatives and multivariate terms. To deal
with these intricate field equations, we employ a particular model
of this theory \cite{31}. In comparison with the Einstein-Hilbert
action, this model can better describe gravitational interactions.
The $f(\mathbb{Q},\mathcal{L}_{m})$ theory takes into account a
nontrivial effects that arise from the energy-momentum distribution
in anisotropic celestial objects, where extreme pressures and
densities prevail. The chosen model is given as
\begin{equation}\label{35a}
f(\mathbb{Q},\mathcal{L}_{m})= 2\mathcal{L}_{m}-\alpha
\mathbb{Q}+\eta,
\end{equation}
where $\alpha,\eta$ are arbitrary constants. The linear model is the
minimal and physically transparent extension of symmetric
teleparallel gravity that includes matter-geometry coupling. The
coupling term $\alpha$ controls how strongly non-metricity affects
gravitational dynamics, directly impacting mass, radius, redshift
and stability. While, the term $\eta$ acts as a background shift,
influencing compactness, pressure balance and allowable stellar
configurations. Both these parameters are constrained using
astrophysical data from compact stars, gravitational waves and
stability analysis. The field equations corresponding to this model
become
\begin{eqnarray}\label{36a}
\varrho &=&\frac{2\alpha(1-e^{-\iota})+r^2 \eta -2r
e^{-\iota}(\alpha \iota'-2\nu')}{2r^2},
\\\label{37a}
p_{r}&=&\frac{2\alpha(1-e^{-\iota})-r^2 \eta+2 r
\nu'e^{-\iota}}{2r^2},
\\\label{38a}
p_{t}&=&\frac{e^{-\iota} (2 \alpha(\iota'-\nu')-r\alpha(\nu')^2
+\alpha\nu' r\iota'-2\alpha r \nu'')-2r\eta}{4 r}.
\end{eqnarray}

In our analysis, the chosen parameter values are constrained by the
regularity of the metric potentials and matter variables. Only
combinations of $\alpha=-0.997$ and $\eta=0.00001$ yields finite
density, positive pressures. The considered parameter values are
physically motivated and systematically constrained through
theoretical limits ensuring reduction to GR, regularity and energy
condition checks as well as observational compatibility with
realistic CSs. The small deviations from unity provide stable and
well-behaved stellar configurations, making them the most suitable
numerical choices for graphical representation. The adopted linear
model in our analysis represents one of the simplest yet physically
meaningful extensions of symmetric teleparallel gravity. This linear
choice allows us to retain analytical tractability while capturing
deviations from GR through the coupling constant $\alpha$ and the
cosmological-like term $\eta$. In the limit $\alpha=1$ and $\eta=0$,
the model reduces to the standard symmetric teleparallel equivalent
to GR. Therefore, the GR limit is smoothly recovered, and the
corresponding metric potentials naturally revert to those obtained
from Einstein's field equations for the same matter configuration.

To study cosmological models and specific properties of spacetime,
we consider a well-known Krori-Barua solution \cite{39}. This
solution provides a robust mathematical phenomenon for analyzing the
behavior of celestial objects under extreme gravitational pressure.
Ultimately, this condition represents an indispensable tool for
assessing the stability of astrophysical structures, offering
critical insights into their long-term evolution and dynamical
behavior. By examining this condition, astronomers determine either
such objects will maintain structural integrity over time or prone
to instabilities, potentially leading to phenomena such as
gravitational collapse or astrophysical disruption. The chosen
Krori-Barua solution with unknown constants $[x,y,z]$ is defined as
\begin{eqnarray}\label{38}
e^{\nu(r)}=e^{y r^{2}+z}, \quad e^{\iota(r)}=e^{x r^{2}}.
\end{eqnarray}
The unknown constants can be found using the first Darmois junction
condition that matches interior and exterior spacetimes at the
hypersurface.

We use the Darmois junction conditions to determine these unknown
parameters. The Darmois conditions ensure the physical and
mathematical consistency when matching two different spacetime
solutions at a common boundary known as a junction surface. These
constraints are used in a variety of contexts including
astrophysical models (modeling stars where an interior solution must
be matched with an exterior solution), wormholes (connecting two
distinct space-time regions with a throat in-between) and
cosmological models (involving different phases or regions of the
universe). These conditions determine how distinct regions can be
smoothly connected at a boundary of spacetime. For matching
condition, we consider exterior region as Schwarzschild spacetime
and interior region as static spherically symmetric spacetime for
the considered CSs.

The outer geometry of the celestial objects is given as
\begin{eqnarray}\label{44a}
ds^{2}_{+}=dt^{2}(1-\frac{2m}{r})-dr^{2}(1-\frac{2m}{r})^{-1}-d\theta^{2}r^{2}
-d\phi^{2}r^{2}\sin^{2}\theta.
\end{eqnarray}
At the surface boundary $(r=\mathbb{R})$ gives
\begin{eqnarray}\label{45a}
g_{tt}&=& e^{y \mathbb{R}^{2}+z}=1-\frac{2m}{\mathbb{R}},
\\\label{46a}
g_{rr}&=& e^{x \mathbb{R}^{2}}= (1-\frac{2m}{\mathbb{R}})^{-1},
\\\label{47a}
g_{tt,r}&=& y \mathbb{R} e^{y \mathbb{R}^{2}+z}=\frac{m}{\mathbb{R}^{2}}.
\end{eqnarray}
Solving the above equations, we obtain
\begin{eqnarray}\label{40}
x=-\frac{1}{\mathbb{R}^2}\ln(1-\frac{2m}{\mathbb{R}}), \quad y=\frac{m}{\mathbb{R}^2 (\mathbb{R}-2m)},
\quad z=\frac{m}{2m-\mathbb{R}}\ln(1-\frac{2m}{\mathbb{R}}).
\end{eqnarray}
\begin{figure}
\epsfig{file=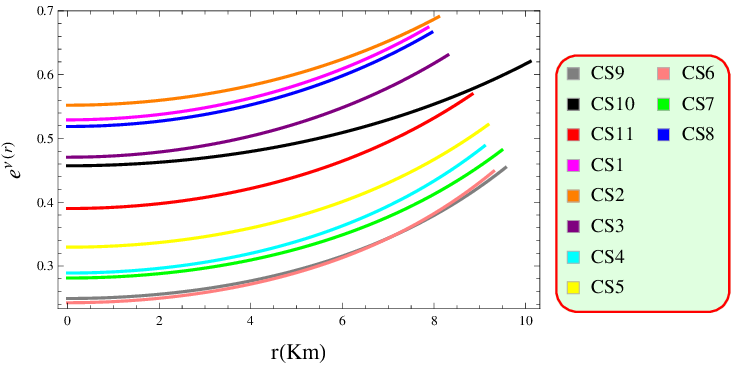,width=.5\linewidth}
\epsfig{file=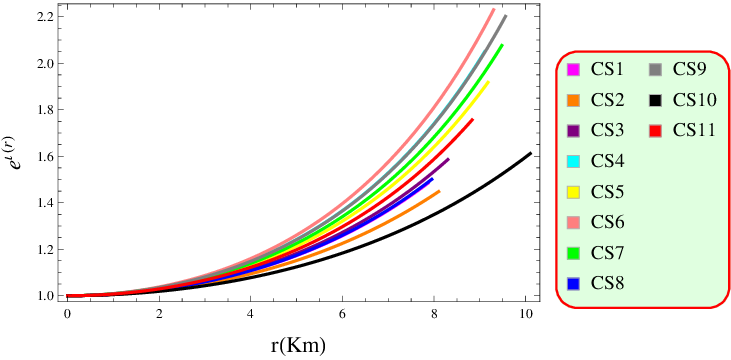,width=.5\linewidth}\caption{Analysis of metric
potential versus radial coordinate.}
\end{figure}

The Krori-Barua metric potentials serve as important indicators for
analyzing singularities and event horizons in relativistic systems.
This solution exhibits non-singular behavior, with metric elements
demonstrating a strictly positive and monotonically increasing trend
throughout the entire spacetime domain. As illustrated in Figure
\textbf{1}, the graphical representation of metric potentials
verifies both their regularity and the required positive, increasing
nature of both metric components. The Krori-Barua solution provides
insights into physical properties, structure and energy distribution
of the celestial objects. In our paper, the measured values of mass
and radius for the celestial objects are compiled in Table
\textbf{1} \cite{40}-\cite{45}. The values of unknown constants
derived from these stellar parameters are presented in Table
\textbf{2}.
\begin{table}\caption{Approximate values of mass and radius of compact stars.}
\begin{center}
\begin{tabular}{|c|c|c|}
\hline Celestial Objects & $m(m_{\odot})$ & $\mathbb{R}(km)$
\\
\hline \textbf{Vela X-1} & 1.77 $\pm$ 0.08 & 9.56 $\pm$ 0.08
\\
\hline \textbf{Cen X-3 } & 1.49 $\pm$ 0.08 & 9.178 $\pm$ 0.13
\\
\hline \textbf{4U 1608-52} & 1.74 $\pm$ 0.01 & 9.3 $\pm$ 0.10
\\
\hline \textbf{PSR J1903+327} & 1.667 $\pm$ 0.021 & 9.48 $\pm$ 0.03
\\
\hline \textbf{4U 1538-52} & 0.87 $\pm$ 0.07 & 7.866 $\pm$ 0.21
\\
\hline \textbf{SAX J1808.4-3658} & 0.9 $\pm$ 0.3 & 7.951 $\pm$ 1.0
\\
\hline \textbf{Her X-1} & 0.85 $\pm$ 0.15 & 8.1 $\pm$ 0.41
\\
\hline \textbf{4U 1820-30} & 1.58 $\pm$ 0.06 & 9.1 $\pm$ 0.4
\\
\hline \textbf{SMC X-4} & 1.29 $\pm$ 0.05 & 8.831 $\pm$ 0.09
\\
\hline \textbf{EXO 1785-248} & 1.30 $\pm$ 0.2 & 10.10 $\pm$ 0.44
\\
\hline \textbf{LMC X-4} & 1.04 $\pm$ 0.09 & 8.301 $\pm$ 0.2
\\
\hline
\end{tabular}
\end{center}
\end{table}
\begin{table}\caption{Approximate values of unknown constants corresponding
to mass-radius of compact stars.}
\begin{center}
\begin{tabular}{|c|c|c|c|}
\hline Celestial Objects & x $(km)^{-2}$ & y $(km)^{-2}$ & z
\\
\hline \textbf{Vela X-1} & 0.00863565 & 0.00657448 & -1.39011
\\
\hline \textbf{Cen X-3 } & 0.00773096 & 0.00544831 & -1.11017
\\
\hline \textbf{4U 1608-52} & 0.00927255 & 0.00711041 & -1.41696
\\
\hline \textbf{4U 1820-30} & 0.00865999 & 0.00633108 & -1.24141
\\
\hline \textbf{PSR J1903+327} & 0.00812966 & 0.0059884 & -1.26880
\\
\hline \textbf{4U 1538-52} & 0.00637763 & 0.00390959 & -0.636511
\\
\hline \textbf{SAX J1808.4-3658} & 0.00642228 & 0.00396096 &
-0.656412
\\
\hline \textbf{Her X-1} & 0.00564142 & 0.00341357 & -0.594098
\\
\hline \textbf{SMC X-4} & 0.00722214 & 0.00484915 &-0.941398
\\
\hline \textbf{EXO 1785-248} & 0.0046774 & 0.00299707 & -0.782873
\\
\hline \textbf{LMC X-4} & 0.00669013 & 0.00424959 & -0.753818
\\
\hline
\end{tabular}
\end{center}
\end{table}

\section{Viable Features of Compact Stars}

Through comprehensive evaluation and graphical analysis, we
systematically examine the physical viability of celestial objects.
Our investigation encompasses multiple key parameters including
density, pressure components, energy conditions, state parameters,
mass-radius relationship, compactness, surface redshift and
stability criteria through adiabatic index and sound speed analysis.
These characteristics are rigorously assessed through detailed
graphical representations in order to validate the physical
acceptability of the celestial objects.

\subsection{Matter Contents}

The investigation of matter contents within celestial objects plays
a fundamental role in elucidating their internal structure and
evolutionary dynamics. These thermodynamic variables particularly
energy density and pressure components naturally attain their
maximum values at the stellar core, a direct consequence of the
extreme density profile characterizing compact astrophysical
objects. The positive-definite nature of these quantities provides
crucial opposition to gravitational compression, thereby maintaining
hydrostatic equilibrium and preventing gravitational collapse. Using
Eqs.(\ref{36a})-(\ref{38a}), we obtain the corresponding field
equations as
\begin{eqnarray}\label{49a}
\varrho&=&\frac{e^{-x r} (8 y r^2 - 2 (1 + x r) \alpha) +2 \alpha -
r^2 \eta}{2 r^2},
\\\label{50a}
p_{r}&=&\frac{4r^2 y e^{-x r} +2 (\alpha -e^{-x r} \alpha) - r^2
\eta}{2r^2},
\\\label{51a}
p_{t}&=& \frac{e^{-x r}((x + x y r^2 - 2 y r (2 + y r^2)) \alpha )-
r \eta}{2 r}.
\end{eqnarray}
Figure \textbf{2} presents a comprehensive analysis of the matter
variables. The plots reveal several key features, as all matter
variables show peak values at stellar center and monotonic
decreasing behavior with increasing radial coordinate, confirming
physical consistency. Figure \textbf{3} shows the rate of change for
the matter variables in each compact stars. These graphs show the
rate of change of matter contents are zero and negative at the
center satisfying the viability of compact stars.
\begin{figure}
\epsfig{file=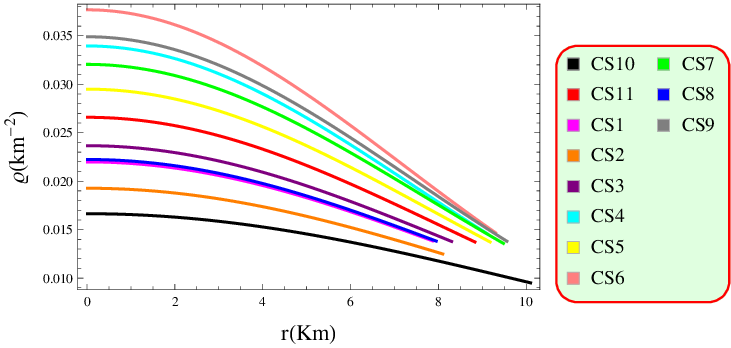,width=.5\linewidth}
\epsfig{file=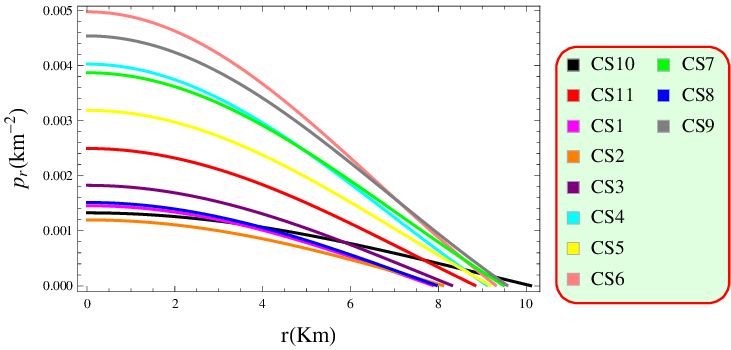,width=.5\linewidth}\center
\epsfig{file=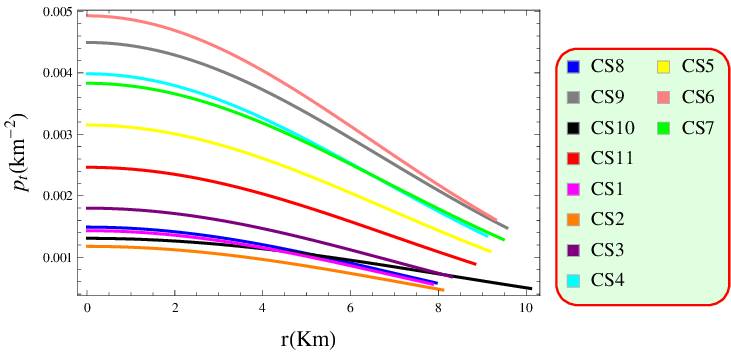,width=.5\linewidth}\caption{Analysis of matter
contents versus radial coordinate.}
\end{figure}
\begin{figure}
\epsfig{file=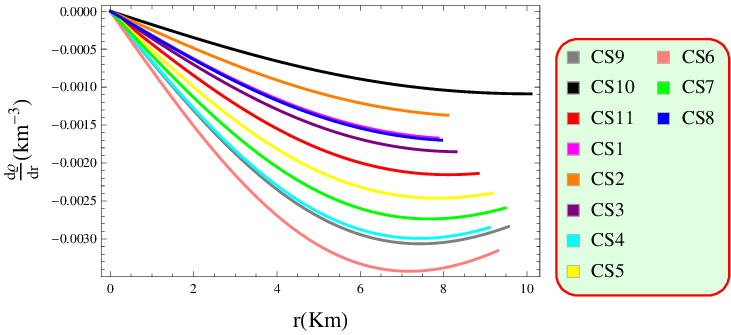,width=.5\linewidth}
\epsfig{file=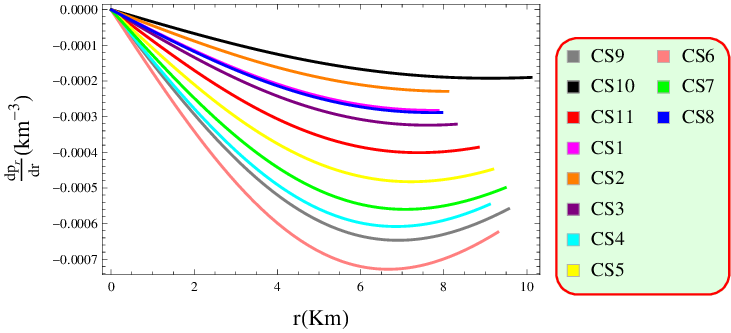,width=.5\linewidth}\center
\epsfig{file=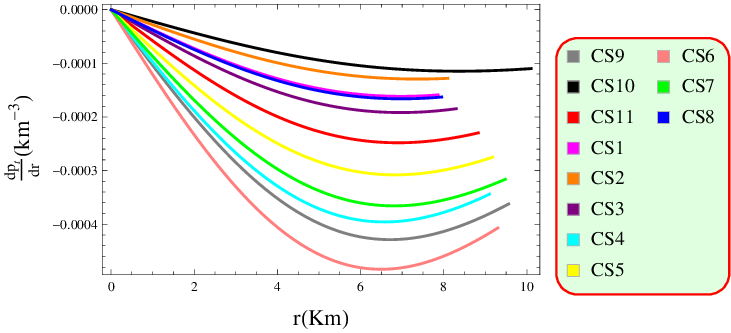,width=.5\linewidth}\caption{Analysis of matter
content's gradient versus radial coordinate.}
\end{figure}

The difference of radial and tangential pressure is known as
anisotropy $(\Delta=p_t-p_r)$ that describes a physical state where
the internal pressure of a celestial body demonstrates directional
dependence. This contrasts with isotropic systems where pressure
remains identical in all orientations. The pressure anisotropy
becomes particularly relevant in extreme environments like neutron
stars and quark stars, where nuclear pressure exhibits directional
behavior \cite{46}. The positive anisotropy creates a repulsive
force $(\Delta>0)$, while negative anisotropy $(\Delta<0)$ manifests
the attractive force in the compact objects. The behavior of change
in pressure components $(\Delta=p_t-p_r)$ is shown in Figures
\textbf{4}. A positive change in pressure components corresponds to
outward-directed repulsive force within all CSs. All these features
collectively establish the existence of physically viable compact
stars within the framework of $f(\mathbb{Q},\mathcal{L}_{m})$
gravity. The consistent behavior across all analyzed cases
demonstrate the robustness of our solution.
\begin{figure}\center
\epsfig{file=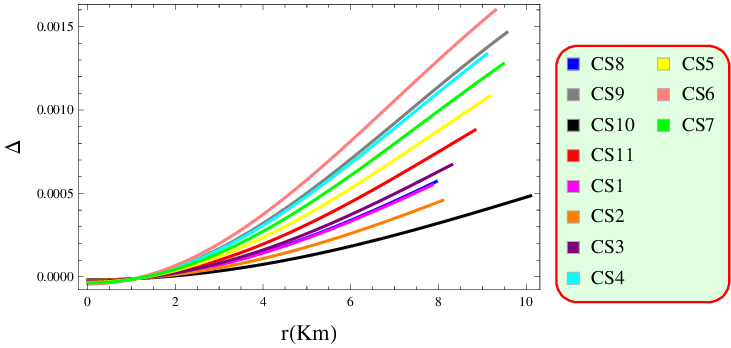,width=.5\linewidth}\caption{Analysis of change
in pressure components versus radial coordinate.}
\end{figure}

The measure of positive anisotropy plays a vital stabilizing role
and prevent the catastrophic collapse of large-scale astrophysical
structures. This force is a key mechanism for supporting massive CSs
against their own immense gravity, preventing them from collapsing
into a black hole. The positive value of anisotropy has profound
implications on CSs, as affects their observable properties like
mass, radius and surface redshift. Positive anisotropy implies that
the pressure along the radial direction is stronger than along the
tangential direction. This reduce the effective attraction inside
the star, making it more prone to collapse. Compact stars with
positive anisotropy are generally more compact for the same mass.

\subsection{Normal Matter}

For physically feasible celestial objects, certain energy
constraints (EC) must be satisfied. These conditions are expressed
as inequalities that place limits on components of the matter-energy
tensor, which govern how matter and energy behave under the action
of gravity. Several types of EC impose distinct limits on energy
density and pressure components, defined as
\begin{itemize}
\item Null EC
\begin{eqnarray}\nonumber
0\leq p_{r}+\varrho, \quad 0\leq p_{t}+\varrho.
\end{eqnarray}
\item Dominant EC
\begin{eqnarray}\nonumber
0\leq \varrho\pm p_{r}, \quad 0\leq \varrho\pm p_{t}.
\end{eqnarray}
\item Weak EC
\begin{eqnarray}\nonumber
0\leq p_{r}+\varrho,\quad 0\leq p_{t}+\varrho, \quad 0\leq \varrho.
\end{eqnarray}
\item Strong EC
\begin{eqnarray}\nonumber
0\leq p_{r}+\varrho, \quad 0\leq p_{t}+\varrho, \quad 0\leq
p_{r}+2p_{t}+\varrho.
\end{eqnarray}
\end{itemize}
Basically, all these EC define nature of matter and energy within
the celestial objects. The viable celestial objects satisfying all
EC, contain the possible normal matter.

The energy conditions impose different limitations on fluid
parameters within the framework of $f(\mathbb{Q},\mathcal{L}_{m})$
gravity. The null energy condition ensures non-negative energy
density for all observers. It ensures that even for light rays, the
energy density is non-negative. It is crucial for the second law of
black hole thermodynamics and for the focusing of light rays by
gravity. Violating this condition is a signature of exotic matter.
The weak energy condition confirms that energy density is
non-negative and flows in a causal manner. The dominant energy
condition guarantees that the speed of energy flow does not exceed
the speed of light. It ensures that the energy flux is time-like or
null. This condition implies the weak EC and supports for physical
plausibility. The strong energy condition (SEC) indicates the
gravitational attraction dominates over repulsion. The fulfillment
of these conditions supports the stability of the CSs, ensuring they
are not prone to collapse or unphysical behavior under the chosen
$f(\mathbb{Q},\mathcal{L}_{m})$ gravity model. This condition is
directly related to the attractive nature of gravity in the
Raychaudhuri equation (which describes the convergence of
geodesics). It is important to note that the SEC is known to be
violated by certain well-established phenomena, most notably the
accelerated expansion of the universe being driven by dark energy.

The satisfaction of the SEC is the classical norm and its violation
is one of the most important discoveries in modern cosmology. The
SEC is significant because it is the direct link between the
energy-momentum content of a spacetime and the attractive nature of
gravity. This is the critical modern context in the realm of
standard and attractive gravitation, which is a cornerstone for
modeling realistic compact stellar objects. This condition ensures
that the energy-momentum content via the Einstein field equation
creates a gravitational field that pulls matter together, rather
than pushing it apart. The SEC appears as a critical term in the
Raychaudhuri equation, which governs the convergence of a bundle of
geodesics (the paths of freely falling particles). Satisfying the
SEC guarantees that the convergence is non-negative, mean nearby
particles are forced to move toward each other over time. The SEC is
a fundamental assumption in the famous Penrose-Hawking singularity
theorems. These theorems state that under SEC, a spacetime
containing a trapped surface (like inside a black hole event horizon
or during the Big Bang) must contain a gravitational singularity.

In the context of compact stars, satisfaction of the SEC during
gravitational collapse ensures that the collapse is irreversible and
relentless. It guarantees that the inward pull of gravity will
overcome any known pressure, leading to the formation of a black
hole singularity. In principle, violation of the SEC could lead to a
stable object like a gravastar. The significance of a satisfied SEC
describes a regime where normal matter dominates. It signifies a
universe where gravity behaves as we classically expect and pulls
things together. The violation of the SEC, signifies the dominance
of an exotic form of energy that causes gravity to become repulsive
on large scales. Figure \textbf{5} signifies normal matter for
considered CSs as all ECs are satisfied in the framework of
$f(\mathbb{Q},\mathcal{L}_{m})$ theory.
\begin{figure}
\epsfig{file=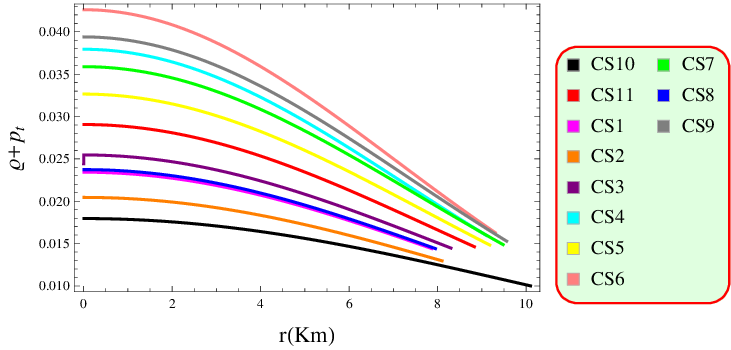,width=.5\linewidth}\epsfig{file=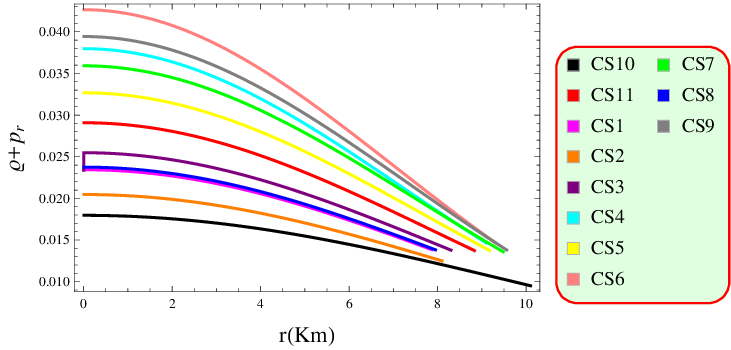,width=.5\linewidth}
\epsfig{file=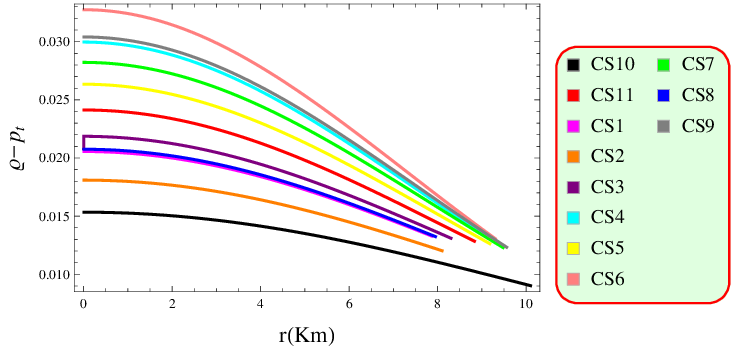,width=.5\linewidth}
\epsfig{file=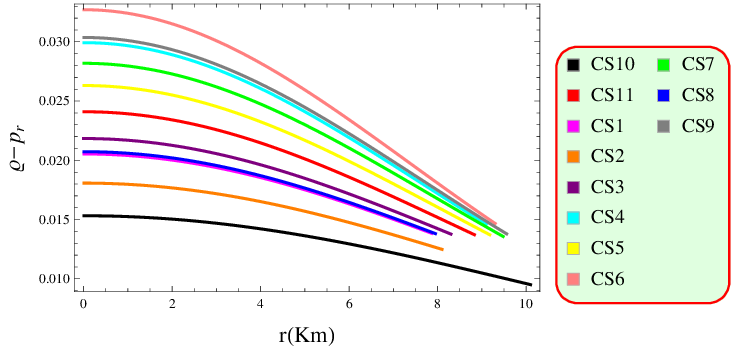,width=.5\linewidth}\caption{Analysis of energy
conditions versus radial coordinate.}
\end{figure}

\subsection{State Parameters}

The state parameters describe relation of matter contents in
physical system. The equation of radial
$(\omega_{r}=\frac{p_{r}}{\varrho})$ and transverse
$(\omega_{t}=\frac{p_{t}}{\varrho})$ state parameters, must satisfy
the interval $[0,1]$ for viable compact objects \cite{47}. Using
Eqs.(\ref{49a})-(\ref{51a}), we have
\begin{eqnarray}\label{52a}
\omega_{r}&=&1-\frac{4 r^2 (y - x \alpha)}{8 y r^2 - 2 (1 + 2 x r^2)
\alpha+ e^{x r^2} (2 \alpha - r^2 \eta)},
\\\label{53a}
\omega_{t}&=& \frac{r^2 (-2 (x - 2 y + (x - y) y r^2) \alpha +
e^{x r^2} \eta)}{-8 y r^2 + 2 (1 + 2 x r^2) \alpha +
e^{x r^2} (r^2 \eta-2 \alpha)}.
\end{eqnarray}
Figure \textbf{6} determines graphical analysis of state parameters
satisfying the required viability of the celestial objects.
\begin{figure}
\epsfig{file=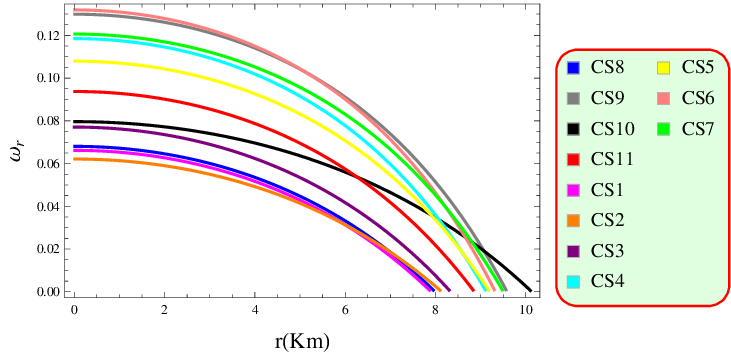,width=.5\linewidth}
\epsfig{file=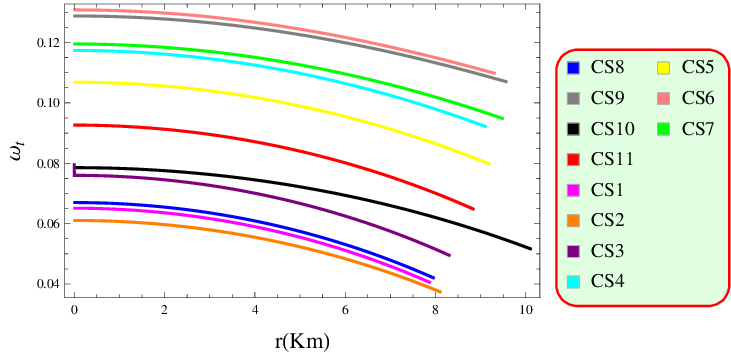,width=.5\linewidth}\caption{Analysis of EoS
parameters versus radial coordinate.}
\end{figure}

\subsection{Physical Features}

The anisotropic celestial objects mass is
\begin{equation}\label{54a}
M=4\pi\int^{\mathbb{R}}_{0} r^{2}\varrho dr.
\end{equation}
The mass distribution behavior in celestial objects based on a
numerical solution with initial condition $M(0)=0$ is shown in
Figure \textbf{7}. The graphic demonstration is consistent with
positive increase in mass and the radial distance. Furthermore, as
radius approaches zero, the mass function converges to zero
indicating regularity at the center of the celestial object. The
structural composition of celestial objects is analyzed by the
compactness function defined as $u=\frac{M}{r}$. Buchdahl \cite{48}
established an important upper limit for the mass to radius ratio,
stating that physically realistic celestial objects must satisfy the
condition $u<4/9$.
\begin{figure}
\epsfig{file=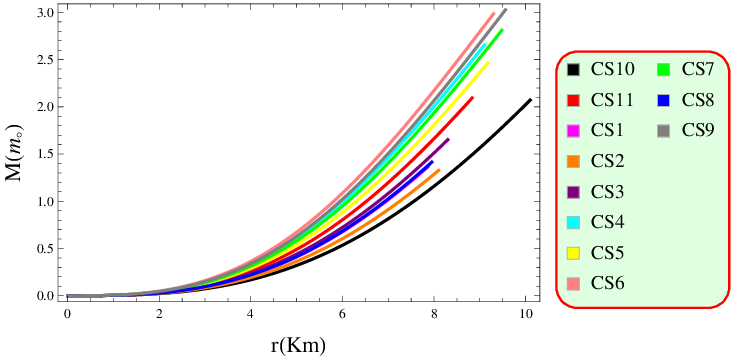,width=.5\linewidth}
\epsfig{file=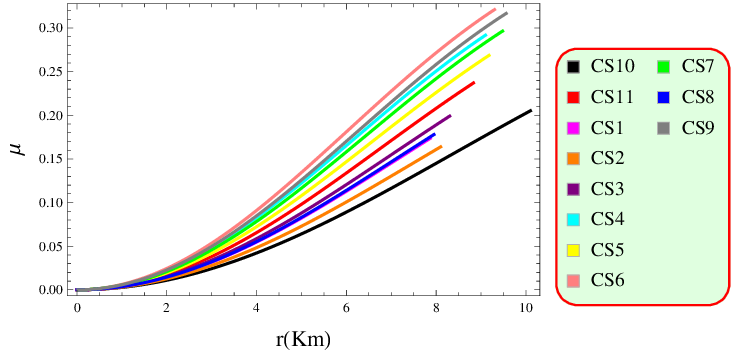,width=.5\linewidth}\center
\epsfig{file=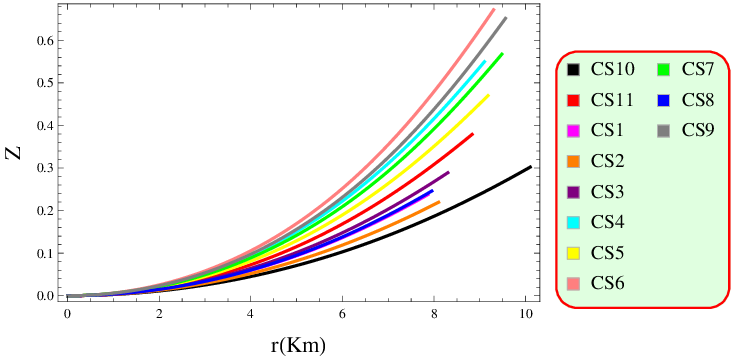,width=.5\linewidth}\caption{Analysis of mass,
compactness and redshift versus radial coordinate.}
\end{figure}

The surface redshift is an important observational metric that
measures the gravitational impact on properties of light emitted by
celestial objects via the relativistic wavelength shift. Additional
information about the object's internal structure and gravitational
field can be gained by expressing this redshift in terms of the
compactness function
\begin{equation}\label{55a}
Z_s = \frac{1}{\sqrt{1-2u}}-1.
\end{equation}
According to a critical condition set by Buchdahl \cite{48}, a
celestial object with a perfect matter distribution must have a
surface redshift of less than 2. In contrast, Ivanov \cite{49} found
that anisotropic configurations had a value of 5.211. As illustrated
in Figure \textbf{7}, the compactness and redshift functions exhibit
increasing trend, asymptotically approaching zero at the center of
the celestial objects. Notably, both functions satisfy their
respective physical constraints ($u < 4/9$ for compactness and
$Z_{s} < 5.211$ for surface redshift), ensuring consistency with
theoretical bounds.

\subsection{Mass-Radius Relation}

The mass-radius $(M-\mathbb{R})$ relation is an essential tool in
the study of CSs. This provides insights into the relationship
between mass, radius and moment of inertia for the CSs. The
mass-radius relationship helps to understand how the mass of a CS
influences its size and structure. The $M-\mathbb{R}$ relation helps
astronomers to compare theoretical models with observational data,
allowing them to constrain the properties of CSs and understand
their internal structure. Several researchers studied
$(M-\mathbb{R})$ relation to provide a fundamental diagnostic for
testing the internal composition of CSs. Tangphati et al studied
$(M-\mathbb{R})$ relation for CSs in the modified theories
\cite{55a,56}. Banerjee et al explored the mass-radius properties of
quark stars formulated with an interacting quark matter within the
framework of Rastall gravity \cite{57}.

Here, we investigate the gravitational mass and radius with the
following condition $e^{\nu}=1-\frac{2m}{r}$. The total mass is
obtained as
\begin{equation}\label{45a}
M= \frac{1}{2}\mathbb{R}(1-c-b \mathbb{R}^2).
\end{equation}
The relationship between the total mass (expressed in $m_{\odot}$)
and the radius (measured in $km$) is presented graphically in Figure
\textbf{8}. Our graphical analysis yields a significant finding, the
identification of the maximum mass exhibited by a CS which we
determine to be 5.047 solar masses. This observation aligns
precisely with the existing literature and the value presented in
Table \textbf{1} of our study. Specifically, this mass value
corresponds to the stellar configuration associated with the CS
designated as Her X-1. Important findings and established data serve
to reinforce the reliability and validity of our research results.
\begin{figure}\center
\epsfig{file=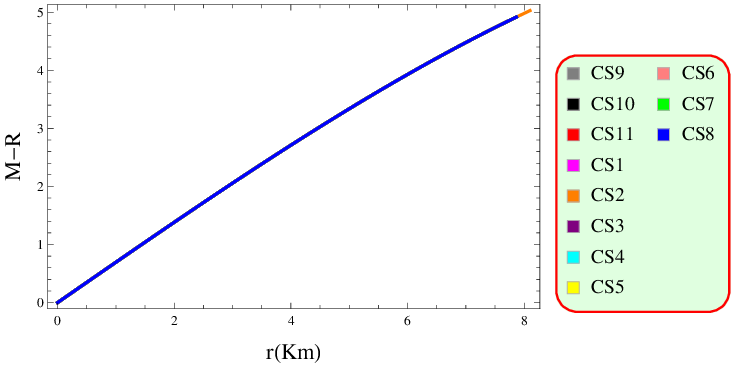,width=.5\linewidth}\caption{Analysis of
mass-radius relation versus radial coordinate.}
\end{figure}

\section{Stability Analysis}

Stability is a key necessarily measure for assessing and exploring
conditions under which the celestial objects remain stable against
several oscillations. Here, we use the sound speed and adiabatic
index methods to analyze the stability of celestial stars. Sound
speed defines the rate at which pressure waves propagate through a
medium, while the adiabatic index characterizes the relationship
between pressure and density changes in the celestial objects.

\subsection{Sound Speed}

The sound speed method reflects speed by which pressure propagates
through a medium. It ensures that no signals travel faster than
light. This criteria rule that the radial and tangential components
of sound speed denoted as $(u_{r}^2=\frac{dp_{r}}{d\varrho})$ and
$(u_{t}^2 =\frac{dp_{t}}{d\varrho})$ along with their difference
$(|u_{t}^2 - u_{r}^2|)$ are confined to the interval $[0,1]$, for
celestial objects to be stable \cite{50,51}. Both components of
sound speed method become
\begin{eqnarray}\label{56a}
u_{r}^2&=&\frac{2\alpha -2 x y r^3 - 2 e^{x r^2} \alpha + x r
\alpha}{-4 x y r^3 + 2 \alpha - 2 e^{x r^2} \alpha+2 x r \alpha +
x^2 r^2 \alpha},
\\\label{57a}
u_{t}^2&=&\frac{r (x + x^2 r - 5 x y r^2 + y (x^2 + 4 y) r^3 -2 x y^2 r^4) \alpha}{
8 x y r^3 + 4 (e^{x r^2}-1) \alpha - 2 x r (2 + x r) \alpha}.
\end{eqnarray}
The stability of the celestial objects is confirmed by the visual
representation of their adherence to the necessary constraints in
Figure \textbf{9}.
\begin{figure}
\epsfig{file=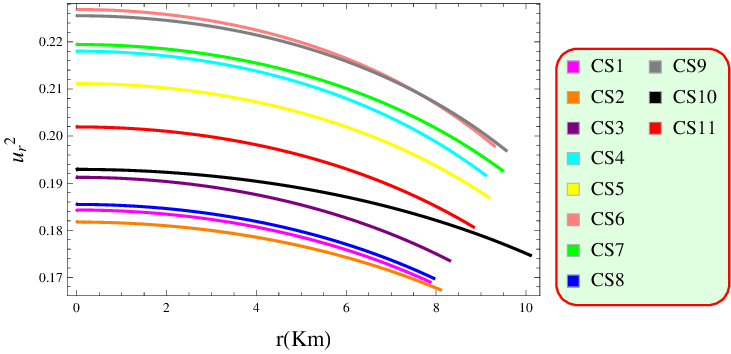,width=.5\linewidth}
\epsfig{file=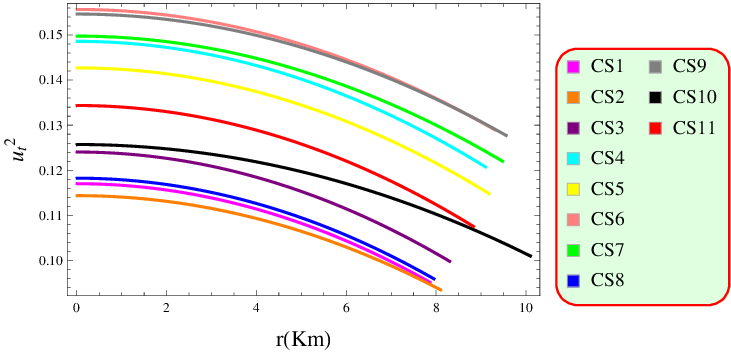,width=.5\linewidth}\center
\epsfig{file=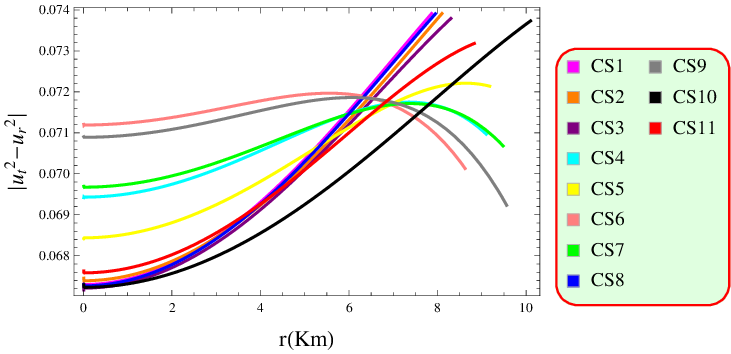,width=.5\linewidth}\caption{Analysis of
stability versus radial coordinate by sound speed method.}
\end{figure}
\begin{figure}
\epsfig{file=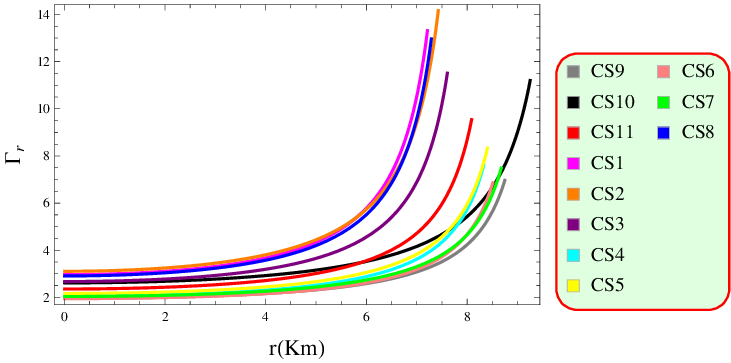,width=.5\linewidth}
\epsfig{file=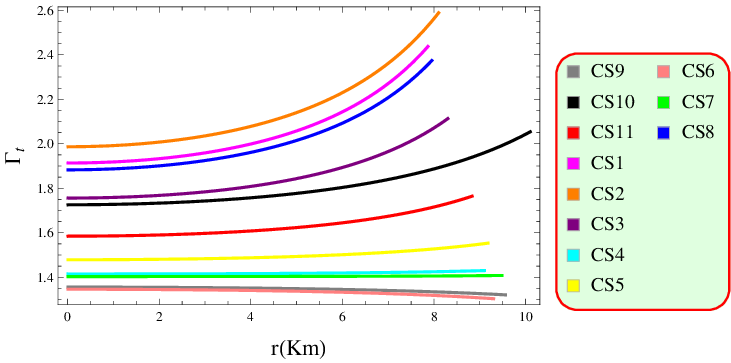,width=.5\linewidth}\caption{Analysis of
stability versus radial coordinate by adiabatic index method.}
\end{figure}

\subsection{Adiabatic Index}

Adiabatic index is another approach to determine the stability of
celestial objects. It measures the stiffness of matter and strongly
resists compression. For viable celestial objects, the radial
$\Gamma_{r}=(\frac{\varrho+p_{r}}{p_{r}}) u_{r}^2$ and tangential
$\Gamma_{t}=(\frac{\varrho+p_{t}}{p_{t}}) u_{t}^2$ components of
adiabatic index must be greater than a critical value $4/3$. A value
below this limit indicates instability and leads to gravitational
collapse \cite{52}. Both components of adiabatic index yield
\begin{eqnarray}\nonumber
\Gamma_{r}&=&\bigg[2 ((e^{x r^2}-1) \alpha + x (2 y r^4 -r^2
\alpha)) ( 2 (1 + x r^2) \alpha -6 y r^2+e^{x r^2}(-2
\\\nonumber
&\times&\alpha+ r^2 \eta))\bigg]\bigg[(\alpha - e^{x r^2}\alpha +  x
r^2 (\alpha + 2 r^2 (x \alpha-2 y))) (4 y r^2 - 2 \alpha
\\\label{58a}
&+& e^{x r^2}(2 \alpha-r^2 \eta))\bigg]^{-1},
\\\nonumber
\Gamma_{t}&=&\bigg[-(2 r^2 (x^2 - 3 x y + y^2 + x (x - y)y r^2)
\alpha ((e^{x r^2}-1) \alpha - r^2 (x \alpha + y
\\\nonumber
&\times&(-4 + (2 + (-x + y) r^2) \alpha) + e^{x
r^2}\eta)))\bigg]\bigg[(\alpha - e^{x r^2} \alpha +   x r^2 (\alpha
+ 2 r^2
\\\label{59a}
&\times&(x \alpha-2 y))) (2 (x -
2 y + (x - y) y r^2) \alpha - e^{x r^2} \eta)\bigg]^{-1}.
\end{eqnarray}
The results presented in Figure \textbf{10}, demonstrate that our
system fulfills this stability requirement $\Gamma > 4/3$.

\section{Conclusions}

Recently developed $f(\mathbb{Q},\mathcal{L}_{m})$ theory attempts
to address a number of unanswered issues in cosmology and
astrophysics. Celestial spheres are fundamental objects in the
universe, discern unique behavior within this modified framework as
compared to general relativity. By comparing theoretical predictions
with observations of celestial objects, we can constrain free
parameters that deviate from general relativity. As a result,
valuable insights can be gained into celestial properties. Our
research investigates structure of celestial objects in this theory
to unravel mysteries of the universe. We have explored the effects
of this theory under extreme curvature and density conditions. To
determine unknown constants in our model, we have employed Darmois
conditions at the boundary of the celestial objects. We have
analyzed stable state of the celestial objects through sound speed
method and adiabatic index criteria. The summary of our findings can
be given as follows.
\begin{itemize}
\item
Both consistent and non-singular metric functions (Figure
\textbf{1}) ensure the smooth spacetime, free from singularities.
\item
The observed matter and pressure profiles are consistent with
theoretical expectations for compact relativistic objects (Figure
\textbf{2}). The radial pressure satisfies the boundary condition,
required for a physically realistic stellar model.
\item
A negative gradient of the matter contents indicate a dense profile
of the suggested stellar objects (Figure \textbf{3}).
\item
The vanishing anisotropy at center satisfies the necessary
conditions for stable anisotropic stellar configurations (Figure
\textbf{4}).
\item
All EC (Figure \textbf{5}) are strictly positive, confirming
ordinary matter within the stellar celestial objects, a fundamental
and necessary requirement for physically viable celestial objects.
\item
The state parameters (Figure \textbf{6}) strictly satisfy
$0<\omega_{r},\omega_{t} < 1$, demonstrating consistency of the
proposed model with realistic matter configurations.
\item
We have found regular mass function at the center of celestial
objects, that monotonically increases with radius (Figure
\textbf{7}).
\item
The compactness factor remains below the Buchdahl limit 4/9, while
the surface redshift does not exceed Ivanove limit 5.2, ensuring the
existence of stable and observationally permissible compact objects.
(Figure \textbf{7}).
\item
The stability criteria evaluated by both methods are fully
satisfied, verifying that the configurations remain stable against
radial perturbations (Figures \textbf{9}-\textbf{10}).
\end{itemize}

Here to note that, Salako \cite{53} examined the LMC X-4 system
exclusively through conformal Killing vectors and established
stability criteria through various physical parameters. Similarly,
Das \cite{53b} derived solutions using observational data from the
4U 1608-52 system. In contrast, our work employs the novel framework
of $f(\mathbb{Q},\mathcal{L}_{m})$ gravity to represent a
significant theoretical departure that enables comprehensive
analysis of modified gravity effects on stellar viability and
stability. Through distinct methodological approaches, we have
systematically examined multiple anisotropic compact objects within
this framework, assessing their structural properties under
gravitational dynamics. Our analysis encompasses the full spectrum
of critical physical parameters, i.e., metric potential, matter
variables, equation of state parameters, energy constraints,
mass-radius relation, compactness, surface redshift and stability.
All parameters satisfy their respective physical constraints,
confirming both viability and stability within this modified
framework.

We have checked whether our system remains in stable state or not in
the presence of $f(\mathbb{Q},\mathcal{L}_{m})$ terms. It is found
that in the presence of modified terms our system remains in a
stable state. Our investigation into the physical aspects has
revealed a dense profile of CSs. We have analyzed various essential
physical parameters such as metric potentials, effective matter
variables, state parameters, energy conditions, mass, compactness,
redshift function and sound speed, which characterize the stellar
system. It is worthwhile to mention here that all these parameters
meet the necessary conditions, indicating the presence of viable and
stable CSs in this modified framework. Furthermore, the chosen
factors for assessing the feasibility and stability of the solution
have satisfied their specified limits. Notably, we observed that all
parameters reach their maximum values when compared to GR
\cite{54a}-\cite{54c} and other modified gravitational theories
\cite{54d,54e}. In the realm of $f(R)$ theory, the results indicate
the instability of the CS Her X-1 associated with the second gravity
model due to the limited range satisfied by the physical quantities
\cite{54f}.

Furthermore, in the framework of $f(R,\mathbb{T}^{2})$ theory, it is
found that CSs are neither physically viable nor stable at the
center \cite{55}. Notably, our solutions demonstrate enhanced
parameter range compared to $f(R)$ gravity models and
$f(R,\mathbb{T}^{2})$ gravity. These results substantiate robust
theoretical construct for modeling extreme gravitational systems
while maintaining consistency with fundamental physical requirements
in this theory. Our findings reveal the feasibility of the proposed
CSs even in the presence of modified terms, underscoring the
robustness of our theoretical framework. In light of these findings,
it can be concluded that all considered CSs exhibit both physical
viability and stability at the center in this modified theory.
Consequently, our results suggest that viable and stable CSs can
exist in this modified theory. Therefore, we conclude that the
solutions we have obtained are physically valid, providing stable
and viable.

Finally, we compare our results with GR. Table \textbf{3} presents
the quantitative comparison between the predictions of the
$f(\mathbb{Q},\mathcal{L}_{m})$ gravity model and those of GR for
the same compact star configurations. This table reports the
deviations from GR and highlights the regimes in which the effects
of $f(\mathbb{Q},\mathcal{L}_{m})$ gravity become significant. In
this table, the central density, surface density, central pressure
and central adiabatic index in the $f(\mathbb{Q},\mathcal{L}_{m})$
gravity framework are denoted by $(A=\varrho_c (10^{15})\text{g
cm}^{-3}),~ (B= \varrho_s (10^{15})\text{g cm}^{-3})$, $(C=p_c
(10^{35})\text{dyne cm}^{-2})$ and $D=\Gamma_r$. In GR, the
corresponding quantities are represented as $(E=\varrho_c
(10^{15})\text{g cm}^{-3} ),~ (F=\varrho_s (10^{15})\text{g
cm}^{-3})$ and $(G=p_c (10^{35})\text{dyne cm}^{-2})$.
\begin{table}\caption{Quantitative Analysis of the $f(\mathbb{Q},\mathcal{L}_{m})$ gravity and
General Relativity.}
\begin{tabular}{|c|c|c|c|c|c|c|c|c|}
\hline Celestial Objects&A&B&C&D&E&F&G\\
\hline Vela X-1& 46.717&18.542&54.686&1.98&46.62&18.37&54.1\\
\hline Cen X-3& 39.519&18.395&38.393&2.17&39.38&18.2&38.2\\
\hline 4U 1608-52&50.437&19.706&59.891&1.97&50.46&19.8&59.4\\
\hline PSR J1903+327 &42.998&18.235&46.434&2.02&42.8&18.3&46.29\\
\hline 4U 1538-52 & 29.459&18.369&17.555&2.99&29.3&18.27&17.2\\
\hline SAX J1808.4-3658& 29.847&18.54&18.241&2.89&29.8&18.5&18.2\\
\hline Her X-1& 25.820&16.669&14.393&3.13&26.2&16.71&14.1\\
\hline 4U 1820-30 &45.487&19.559&48.686&2.99&45.5&19.7&48\\
\hline SMC X-4& 35.653&18.315&30.121&2.35&35.55&18.37&29.7\\
\hline EXO 1785-248 &22.409&12.722&15.96&2.61&22.28&12.7&15.8\\
\hline LMC X-4& 31.774&18.46&22.088&2.71&31.72&18.4&21.6\\
\hline
\end{tabular}
\end{table}

\section*{Appendix A: Calculation of $\mathbb{Q}$}
\renewcommand{\theequation}{A\arabic{equation}}
\setcounter{equation}{0}

\begin{eqnarray}\label{A1}
\mathbb{Q} &\equiv&
-g^{\chi\tau}\big(\mathcal{L}^{\gamma}_{~\upsilon\chi}
\mathcal{L}^{\upsilon}_{~\tau\gamma}-
\mathcal{L}^{\gamma}_{~\upsilon\gamma}\mathcal{L}^{\upsilon}_{~\chi\tau}\big),
\\\label{A2}
\mathcal{L}^{\gamma}_{~\upsilon\chi}&=&-\frac{1}{2}g^{\gamma\varsigma}
\big(\mathbb{Q}_{\chi\upsilon\varsigma}+
\mathbb{Q}_{\upsilon\varsigma\chi}-\mathbb{Q}_{\varsigma\upsilon\chi}\big),
\\\label{A3}
\mathcal{L}^{\upsilon}_{~\tau\gamma}&=&-\frac{1}{2}g^{\upsilon\psi}
\big(\mathbb{Q}_{\gamma\tau\psi}+
\mathbb{Q}_{\tau\gamma\psi}-\mathbb{Q}_{\psi\tau\gamma}\big),
\\\label{A4}
\mathcal{L}^{\gamma}_{~\upsilon\gamma}&=&-\frac{1}{2}g^{\gamma \psi}
\big(\mathbb{Q}_{\gamma\upsilon\psi}+
\mathbb{Q}_{\upsilon\psi\gamma}-\mathbb{Q}_{\psi\gamma\upsilon}\big),
\\\label{A5}
\mathcal{L}^{\upsilon}_{~\chi\tau}&=&-\frac{1}{2}g^{\upsilon\psi}
\big(\mathbb{Q}_{\tau\chi\psi}+
\mathbb{Q}_{\chi\psi\tau}-\mathbb{Q}_{\psi\chi\tau} \big).
\end{eqnarray}
Therefore, we get
\begin{eqnarray}\label{A6}
-g^{\chi\tau}\mathcal{L}^{\gamma}_{~\upsilon\chi}\mathcal{L}^{\upsilon}_{~\tau\gamma}
&=&-\frac{1}{4}\big(2\mathbb{Q}^{\gamma\tau\psi}
\mathbb{Q}_{\psi\gamma\tau}-\mathbb{Q}^{\gamma\tau\psi}
\mathbb{Q}_{\gamma\tau\psi}\big),
\\\label{A7}
g^{\chi\tau}\mathcal{L}^{\gamma}_{~\upsilon\gamma}\mathcal{L}^{\upsilon}_{~\chi\tau}
&=&\frac{1}{4}g^{\chi\tau}g^{\upsilon\varsigma}\mathbb{Q}_\upsilon
\big(\mathbb{Q}_{\tau\chi\varsigma}+\mathbb{Q}_{\chi\varsigma\tau}
-\mathbb{Q}_{\varsigma\tau\chi}\big),
\\\label{A8}
\mathbb{Q}&=&-\frac{1}{4}\big(2\mathbb{Q}^{\gamma\tau\chi}\mathbb{Q}_{\chi\gamma\tau}
-\mathbb{Q}^{\gamma\tau\chi}
\mathbb{Q}_{\gamma\tau\chi}-2\mathbb{Q}^{\gamma}\tilde{\mathbb{Q}}_{\gamma}
+\mathbb{Q}^{\gamma}\mathbb{Q}_{\gamma}\big).
\end{eqnarray}
Applying Eq.(\ref{22a}), we obtain
\begin{eqnarray}\nonumber
\mathcal{P}^{\gamma\chi\tau}&=&\frac{1}{4}\bigg[-\mathbb{Q}^{\gamma\chi\tau}
+\mathbb{Q}^{\chi\gamma\tau}+\mathbb{Q}^{\tau\gamma\chi}
+\mathbb{Q}^{\gamma}g^{\chi\tau}-\tilde{\mathbb{Q}}^{\gamma}
g^{\chi\tau}- \frac{1}{2}(g^{\gamma\chi}\mathbb{Q}^{\tau}
\\\label{A9}
&+&g^{\gamma\tau}\mathbb{Q}^{\chi})\bigg],
\\\label{A10}
-\mathbb{Q}_{\gamma\chi\tau}\mathcal{P}^{\gamma\chi\tau}
&=&-\frac{1}{4}\big(-\mathbb{Q}^{\gamma\chi\tau}
\mathbb{Q}_{\gamma\chi\tau}+2\mathbb{Q}_{\gamma\chi\tau}
\mathbb{Q}^{\chi\gamma\tau} +\mathbb{Q}^{\gamma}\mathbb{Q}_{\gamma}
-2\mathbb{Q}_{\gamma}\tilde{\mathbb{Q}}^{\gamma}\big)\\\nonumber
&=&\mathbb{Q}.
\end{eqnarray}

\section*{Appendix B: Variation of $\delta \mathbb{Q}$}

\renewcommand{\theequation}{B\arabic{equation}}
\setcounter{equation}{0}

We take non-metricity as
\begin{eqnarray}\label{B1}
\mathbb{Q}_{\gamma\chi\tau}&=&\nabla_{\gamma}g_{\chi\tau},
\\\label{B2}
\mathbb{Q}^{\gamma}_{~\chi\tau}&=&g^{\gamma\upsilon}
\mathbb{Q}_{\upsilon\chi\tau}=
g^{\gamma\upsilon}\nabla_{\upsilon}g_{\chi\tau}=\nabla^{\gamma}
g_{\chi\tau},
\\\label{B3}
\mathbb{Q}^{~~\chi}_{\gamma~~\tau}&=&g^{\chi\upsilon}
\mathbb{Q}_{\gamma\upsilon\tau}=g^{\chi\upsilon}\nabla_{\gamma}
g_{\upsilon\tau}=-g_{\upsilon\tau}\nabla_{\gamma}g^{\chi\upsilon},
\\\label{B4}
\mathbb{Q}^{~~~\tau}_{\gamma\chi}&=&g^{\tau\upsilon}
\mathbb{Q}_{\gamma\chi\upsilon}=g^{\tau\upsilon}\nabla_{\gamma}
g_{\chi\upsilon}=-g_{\chi\upsilon}\nabla_{\gamma}g^{\tau\upsilon},
\\\label{B5}
\mathbb{Q}^{\gamma\chi}_{~~\tau}&=&g^{\gamma\upsilon}
g^{\chi\varsigma}\nabla_{\upsilon}g_{\varsigma\tau}
=g^{\chi\varsigma}\nabla^{\gamma}g_{\varsigma\tau}=-g_{\varsigma\tau}
\nabla^{\gamma}g^{\chi\varsigma},
\\\label{B6}
\mathbb{Q}^{\gamma~\tau}_{~\chi}&=&g^{\gamma\upsilon}
g^{\tau\varsigma}\nabla_{\upsilon}g_{\chi\varsigma}
=g^{\tau\varsigma}\nabla^{\gamma}g_{\chi\varsigma}=-g_{\chi\varsigma}
\nabla^{\gamma}g^{\tau\varsigma},
\\\label{B7}
\mathbb{Q}^{~~\chi\tau}_{\gamma}&=&g^{\chi\varsigma}g^{\tau
\upsilon}\nabla_{\gamma}g_{\varsigma\upsilon}
=-g^{\chi\varsigma}g_{\varsigma\upsilon}\nabla_{\gamma}g^{\tau\upsilon}
=-\nabla_{\gamma}g^{\chi\tau},
\\\label{B8}
\mathbb{Q}^{\gamma\chi\tau}&=&-\nabla^{\gamma}g_{\chi\tau}.
\end{eqnarray}
Also,
\begin{eqnarray}\nonumber
\delta \mathbb{Q}&=&-\frac{1}{4}\delta\bigg(-\mathbb{Q}
^{\gamma\tau\chi} \mathbb{Q}_{\gamma\tau\chi}
+2\mathbb{Q}^{\gamma\tau\chi}\mathbb{Q}_{\chi\gamma\tau}
-2\mathbb{Q}^{\gamma}\tilde{\mathbb{Q}}_{\gamma}
+\mathbb{Q}^{\gamma}\mathbb{Q}_{\gamma}\bigg),
\\\nonumber
&=&-\frac{1}{4}\bigg(-\delta \mathbb{Q}^{\gamma \tau\chi}
\mathbb{Q}_{\gamma\tau\chi} -\mathbb{Q}^{\gamma\tau\chi}\delta
\mathbb{Q}_{\gamma\tau\chi} +2\delta \mathbb{Q}^{\gamma
\tau\chi}\mathbb{Q}_{\chi\gamma\tau}
\\\nonumber
&+&2\mathbb{Q}^{\gamma \tau\chi}\delta \mathbb{Q}_{\chi\gamma\tau}
-2\delta \mathbb{Q}^{\gamma}\tilde{\mathbb{Q}}_{\gamma}
-2\mathbb{Q}^{\gamma}\delta\tilde{\mathbb{Q}}_{\gamma} +\delta
\mathbb{Q}^{\gamma}\mathbb{Q}_{\gamma} +\mathbb{Q}^{\gamma}\delta
\mathbb{Q}_{\gamma}\bigg),
\\\nonumber
&=&-\frac{1}{4}\bigg[\mathbb{Q}_{\gamma\tau\chi}
\nabla^{\gamma}\delta g^{\tau\chi}-\mathbb{Q}^{\gamma \tau\chi}
\nabla_{\gamma}\delta g_{\tau\chi}-2\mathbb{Q}_{\chi\gamma\tau}
\nabla^{\gamma }\delta
g^{\tau\chi}+2\mathbb{Q}^{\gamma\tau\chi}\nabla_{\chi}\delta
g_{\gamma\tau}\\\nonumber
&+&2\tilde{\mathbb{Q}}_{\gamma}\nabla^{\gamma}g^{\chi\tau}\delta
g_{\chi\tau}+2\tilde{\mathbb{Q}}_{\gamma}g_{\chi\tau}\nabla^{\gamma}\delta
g^{\chi\tau}-2\mathbb{Q}^{\gamma}\nabla^{\upsilon}\delta
g_{\gamma\upsilon}-\mathbb{Q}Q_{\gamma}\nabla^{\gamma}g^{\chi\tau}\delta
g_{\chi\tau}\\
\label{B9} &-&\mathbb{Q}_{\gamma}g_{\chi\tau}\nabla^{\gamma}\delta
g^{\chi\tau}-\mathbb{Q}^{\gamma}\nabla_{\gamma}g^{\chi\tau}\delta
g_{\chi\tau}-\mathbb{Q}^{\gamma}g_{\chi\tau}\nabla_{\gamma}\delta
g^{\chi\tau}\bigg],
\end{eqnarray}
\begin{eqnarray}\label{B10}
\delta g_{\chi\tau}&=&-g_{\chi\gamma }\delta
g^{\gamma\upsilon}g_{\upsilon\tau},
\\\label{B11}
-\mathbb{Q}^{\gamma \tau\varsigma}\nabla_{\gamma}\delta
g_{\tau\varsigma}&=&-\mathbb{Q}^{\gamma
\tau\varsigma}\nabla_{\gamma}\big(-g_{\tau\upsilon}\delta
g^{\upsilon\upsilon}g_{\upsilon\varsigma}\big),
\\\label{B12}
2\mathbb{Q}^{\gamma\tau\varsigma}\nabla_{\varsigma}\delta g_{\gamma
\tau}&=&-4\mathbb{Q}^{~~\gamma\varsigma}_{\chi}\mathbb{Q}_{\varsigma\gamma\tau}\delta
g^{\chi\tau}-2\mathbb{Q}_{\tau\varsigma\gamma}\nabla^{\gamma}g^{\tau\varsigma},
\\\nonumber
-2\mathbb{Q}^{\varsigma}\nabla^{\upsilon}\delta
g_{\varsigma\upsilon}&=&2\mathbb{Q}^{\gamma}\mathbb{Q}_{\tau\gamma
\chi}\delta
g^{\chi\tau}+2\mathbb{Q}_{\chi}\tilde{\mathbb{Q}}_{\tau}\delta
g^{\chi\tau}
\\\label{B13}
&+&2\mathbb{Q}_{\tau}g_{\gamma\varsigma}\nabla^{\gamma}g^{\tau\varsigma},
\end{eqnarray}
\begin{equation}\label{B14}
\delta \mathbb{Q}=2\mathcal{P}_{\gamma
\tau\varsigma}\nabla^{\gamma}\delta
g^{\tau\varsigma}-\big(\mathcal{P}_{\chi\gamma\upsilon}\mathbb{Q}^{~~\gamma
\upsilon}_{\tau}
-2\mathbb{Q}^{\gamma\upsilon}_{~~~\chi}\mathcal{P}_{\gamma\upsilon\tau}\big)\delta
g^{\chi\tau},
\end{equation}
where
\begin{eqnarray}\nonumber
2\mathcal{P}_{\gamma\tau\varsigma}&=&-\frac{1}{4}\bigg[2\mathbb{Q}_{\gamma
\tau\varsigma}-
2\mathbb{Q}_{\varsigma\gamma\tau}-2\mathbb{Q}_{\tau\varsigma\gamma}+2
\mathbb{Q}_{\tau}g_{\gamma \varsigma}
\\\label{B15}
&+&2(\tilde{\mathbb{Q}}_{\gamma}-\mathbb{Q}_{\gamma})g
_{\tau\varsigma}\bigg],
\\\nonumber
4\big(\mathcal{P}_{\chi\gamma\upsilon}\mathbb{Q}^{~~\gamma
\upsilon}_{\tau}
-2\mathbb{Q}^{\gamma\upsilon}_{~~~\chi}\mathcal{P}_{\gamma\upsilon\tau}\big)&=&
2\mathbb{Q}^{\gamma \upsilon}_{~~~\tau}\mathbb{Q}
_{\gamma\upsilon\chi}-4 \mathbb{Q}^{~~\gamma
\upsilon}_{\chi}\mathbb{Q}_{\upsilon\gamma\tau}
+2\tilde{\mathbb{Q}}^{\gamma}\mathbb{Q}_{\gamma\chi\tau}
\\\label{B16}
&+&2\mathbb{Q}^{\gamma }\mathbb{Q}_{\tau\gamma\chi}
+2\mathbb{Q}_{\chi}\tilde{\mathbb{Q}}_{\tau}-
\mathbb{Q}^{\gamma}\mathbb{Q}_{\gamma\chi\tau}.
\end{eqnarray}

\section*{Appendix C: Derivation of the Field Equations}
\renewcommand{\theequation}{C\arabic{equation}}
\setcounter{equation}{0}

Here, we show that Eq.(28) leads to Eqs.(31)-(33). The field
equations (28) are
\begin{eqnarray*}\nonumber
\frac{1}{2}f_{\mathcal{L}_{m}}(g_{\chi\tau}\mathcal{L}_{m}-\mathbb{T}_{\chi\tau})&=&
\frac{2}{\sqrt{-g}} \nabla_{\upsilon} (f_{\mathbb{Q}}\sqrt{-g}
\mathcal{P}^{\upsilon}_{~\chi\tau})+ \frac{1}{2}fg_{\chi\tau}
\\\label{29a}
&+&f_{\mathbb{Q}} (\mathcal{P}_{\chi\upsilon\gamma}
\mathbb{Q}_{\tau}^{~\upsilon\gamma}
-2\mathbb{Q}^{\upsilon\gamma}_{~~\chi}
\mathcal{P}_{\upsilon\gamma\tau}),
\end{eqnarray*}
where $f_{\mathcal{L}_{m}}=\frac{\partial f}{\partial
\mathcal{L}_{m}}$ and $f_{\mathbb{Q}}=\frac{\partial f}{\partial
\mathbb{Q}}$. We consider a static spherical spacetime with metric
signatures $(+,-,-,-)$ (29). By the coincident gauge, where
$\Gamma^{\upsilon}_{~\chi\tau}=0$ and $\nabla\rightarrow\delta$, we
get the following quantities
\begin{equation*}\label{30b}
\sqrt{-g}=e^{\frac{\nu+\iota}{2}}r^2
\sin^{2}\theta,\quad\partial_r\ln\sqrt{-g}=\frac{\nu'+\iota'}{2}+\frac{2}{r}.
\end{equation*}
In order to describe matter distribution of spacetime, we consider
matter-Lagrangian as $\mathcal{L}_{m}=p_{r}$. The energy-momentum
tensor describe distribution of matter and energy within a system
and its parameters determine physical properties that govern
dynamics of the system. The choice, $\mathcal{L}_{m}=p_r$, yields a
stress-energy tensor of the form $\mathbb{T}_{\tau\chi}$ that
maintains compatibility with the standard conservation equation in
the GR limit. Moreover, this choice avoids the introduction of extra
dynamical degrees of freedom associated with derivatives of the
energy density. With $\mathcal{L}_{m}=p_r$, the variation
$\frac{\partial \mathcal{L}_{m}}{\partial g^{\chi\tau}}$ depends
only on metric components through the pressure, simplifying the
matter coupling terms in the field equations. As a result, the
modified gravity contributions $(f_{\mathcal{L}_{m}}=\frac{\partial
f}{\partial \mathcal{L}_{m}},~f_{\mathbb{Q}}=\frac{\partial
f}{\partial \mathbb{Q}})$ remain algebraically tractable, enabling
closed-form relations for effective density and pressures.

The four-velocity and four-vector are given as
$\mathbb{U}_{\tau}=(e^{\nu},0,0,0),~
\mathbb{V}_{\tau}=(0,e^{\iota},0,0)$, satisfying
$\mathbb{U}^{\tau}\mathbb{U}_{\tau}=1,~
\mathbb{V}^{\tau}\mathbb{V}_{\tau}=-1,~
\mathbb{U}_{\tau}\mathbb{V}^{\tau}=0.$ Thus Eq.(30) gives
\begin{eqnarray}\nonumber
\mathbb{T}_{00}=e^{\nu}\varrho,~\mathbb{T}_{11}=
e^{\iota}p_r,~\mathbb{T}_{22}=r^2 p_t,~\mathbb{T}_{33}=~r^2
\sin^{2}\theta p_t.
\end{eqnarray}
Using the coincident gauge, we have
\begin{equation*}
\mathbb{Q}_{rtt}=e^{\nu}\nu',~ \mathbb{Q}_{rrr}=-e^{\iota}\iota',~
\mathbb{Q}_{r\theta\theta}=-2r,~
\mathbb{Q}_{r\phi\phi}=-2r^2\sin^{2}\theta,
\end{equation*}
\begin{equation*}
\mathbb{Q}_{r}=\nu'\iota'+\frac{4}{r},~\mathbb{\bar{Q}}_{r}=\iota'.
\end{equation*}
The non-zero components of superpotential are
\begin{equation}\nonumber
\mathcal{P}^{r}_{~00}=\frac{\nu' e^{\nu-\iota}}{4},~
\mathcal{P}^{r}_{~11}=\frac{\iota'}{4},~\mathcal{P}^{r}_{~22}=\frac{-r
e^{-\iota}}{2}.
\end{equation}
The divergence term (only $\upsilon=r$ contributes) gives
\begin{eqnarray}\nonumber
\frac{2}{\sqrt{-g}} \nabla_{\upsilon}(f_{\mathbb{Q}}\sqrt{-g}
\mathcal{P}^{r}_{~\chi\tau})=2(f'_{\mathbb{Q}}
\mathcal{P}^{r}_{~\chi\tau}+f_{\mathbb{Q}}(\mathcal{P}^{r}_{~\chi\tau})'
+f_{\mathbb{Q}}\mathcal{P}^{r}_{~\chi\tau}(\frac{\nu'+\iota'}{2}+\frac{2}{r})).
\end{eqnarray}
For the sake of simplicity, we use the notation
\begin{equation*}\label{23b}
\mathcal{C}_{\chi\tau}=-f_{\mathbb{Q}}(\mathcal{P}_{\chi}^{~\gamma\upsilon}\mathbb{Q}_{\gamma\upsilon\tau}-
2\mathbb{Q}_{\upsilon\gamma\chi}
\mathcal{P}^{\upsilon\gamma}_{~~\tau}).
\end{equation*}
Direct evaluation with nonzero $\mathbb{Q}_{\gamma\chi\tau}$ and
nonzero superpotential gives
\begin{eqnarray}\nonumber
\frac{\mathcal{C}_{00}}{g_{00}}&=&\frac{f_{\mathbb{Q}}}{2
r^{2}e^{\iota}}((2+r \nu')(e^{\iota}-1)+r \iota'(e^{\iota}+1))+
\frac{f_{\mathbb{Q}\mathbb{Q}}\mathbb{Q}'}{ r
e^{\iota}}(e^{\iota}-1),
\\\nonumber
\frac{\mathcal{C}_{11}}{g_{11}}&=&\frac{f_{\mathbb{Q}}}{2
r^{2}e^{\iota}}((e^{\iota}-1)(2+r \nu'+r \iota')-2r\iota')+
\frac{f_{\mathbb{Q}\mathbb{Q}}\mathbb{Q}'}{ r
e^{\iota}}(e^{\iota}-1),
\\\nonumber
\frac{\mathcal{C}_{22}}{g_{22}}&=&\frac{-1}{4 r
e^{\iota}}(f_{\mathbb{Q}}
(\nu^{''}-2r+(r\nu'+2e^{\iota})\iota')+2(e^{\iota}-2)\nu'-r(\nu')^2)-
2f_{\mathbb{Q}\mathbb{Q}}\mathbb{Q}'\nu'.
\end{eqnarray}
Using the above equations, we get
\begin{eqnarray}\nonumber
\frac{1}{2}f_{\mathcal{L}_{m}} (\varrho-
\mathcal{L}_{m})&=&\frac{-f}{2}+\frac{1}{g_{00}}(\frac{-2}{\sqrt{-g}}
\nabla_{\upsilon} (f_{\mathbb{Q}}\sqrt{-g}
\mathcal{P}^{\upsilon}_{~00})+\frac{\mathcal{C}_{00}}{g_{00}},
\\\nonumber \frac{1}{2}f_{\mathcal{L}_{m}} (\varrho-
\mathcal{L}_{m})&=&\frac{1}{2 r^{2}e^{\iota}}\bigg(2r
f_{\mathbb{Q}\mathbb{Q}}\mathbb{Q}'(-1+e^{\iota})+
f_{\mathbb{Q}}((2+r \nu')(e^{\iota}-1)
\\\label{32a}
&+& r\iota'(1+e^{\iota}))\bigg)+\frac{f}{2},
\end{eqnarray}
\begin{eqnarray}\nonumber
\frac{1}{2}f_{\mathcal{L}_{m}} (p_r -
\mathcal{L}_{m})&=&\frac{-f}{2}+\frac{1}{g_{11}}(\frac{-2}{\sqrt{-g}}
\nabla_{\upsilon} (f_{\mathbb{Q}}\sqrt{-g}
\mathcal{P}^{\upsilon}_{~11})+\frac{\mathcal{C}_{11}}{g_{11}},
\\\nonumber
\frac{1}{2}f_{\mathcal{L}_{m}}(p_{r}-\mathcal{L}_{m})&=&\frac{-1}{2
r^{2}e^{\iota}}\bigg(2r
f_{\mathbb{Q}\mathbb{Q}}\mathbb{Q}'(e^{\iota}-1)
+(e^{\iota}-1)f_{\mathbb{Q}}(2+r\nu'+r\iota')
\\\label{33a}
&-&2r\nu')\bigg)-\frac{f}{2},
\end{eqnarray}
\begin{eqnarray}\nonumber
\frac{1}{2}f_{\mathcal{L}_{m}} (p_t -
\mathcal{L}_{m})&=&\frac{-f}{2}+\frac{1}{g_{22}}(\frac{-2}{\sqrt{-g}}
\nabla_{\upsilon} (f_{\mathbb{Q}}\sqrt{-g}
\mathcal{P}^{\upsilon}_{~22})+\frac{\mathcal{C}_{22}}{g_{22}},
\\\nonumber
\frac{1}{2}f_{\mathcal{L}_{m}}(p_{t}-\mathcal{L}_{m})&=&
\frac{-1}{4re^{\iota}}\bigg(-2rf_{\mathbb{Q}\mathbb{Q}}\mathbb{Q}'\nu'+
f_{\mathbb{Q}}(2\nu'(e^{\iota}-2)-r\nu'^{2} +\iota'(2e^{\iota}
\\\label{34a}
&+&r\nu')-2r\nu'')\bigg)-\frac{f}{2}.
\end{eqnarray}
These are the required field equations (31)-(33).\\\\
\textbf{Data Availability Statement:} No new data were generated or
analyzed in support of this research.

\end{document}